\documentclass{aa}  
\usepackage{natbib} 
\usepackage{verbatim} 
\usepackage{graphicx} 
\usepackage{txfonts}
\usepackage{amsmath}
\usepackage{array}

\newcommand\T{\rule{0pt}{2.6ex}}       
\newcommand\B{\rule[-1.2ex]{0pt}{0pt}} 
\usepackage{multirow} 
\usepackage{textcomp} 
\usepackage{float}
\usepackage{bm} 
\usepackage{hyperref}
\newcommand{\um}[1]{\,\mathrm{#1}}
\newcommand{\ump}[2]{\,\mathrm{#1}^{#2}}

\begin{document}

   \title{Coevolution of black hole accretion and star formation in galaxies up to $z=3.5$}


   \author{R. Carraro\inst{\ref{ifa-uv},\ref{unipd} 
                            \thanks{Email: carrarorosamaria@gmail.com}} \and
            G. Rodighiero\inst{\ref{unipd},\ref{oapd}} \and
            P. Cassata\inst{\ref{unipd},\ref{oapd}} \and
            M. Brusa\inst{\ref{unibo},\ref{oabo}} \and
            F. Shankar\inst{\ref{south}} \and
            I. Baronchelli\inst{\ref{unipd}} \and
            E. Daddi\inst{\ref{paris}} \and
            I. Delvecchio\inst{\ref{paris}} \and
            A. Franceschini\inst{\ref{unipd}} \and
            R. Griffiths\inst{\ref{uhawaii},\ref{Carnegie}} \and
            C. Gruppioni\inst{\ref{oabo}} \and
            E. L\'opez-Navas\inst{\ref{ifa-uv}} \and
            C. Mancini\inst{\ref{unipd}} \and
            S. Marchesi\inst{\ref{clemson},\ref{oabo}} \and
            M. Negrello\inst{\ref{cardiff}}  \and
            A. Puglisi\inst{\ref{durham}} \and
            E. Sani\inst{\ref{eso}} \and
            H. Suh\inst{\ref{subaru}}\thanks{Subaru Fellow} 
          }

   \institute{Instituto de F\'\i{}sica y Astronom\'\i{}a, 
                        Universidad de Valpara\'\i{}so,
              Gran Bretaña 1111, Playa Ancha, Valpara\'\i{}so, Chile\label{ifa-uv}
         \and
              $^{}$Dipartimento di Fisica e Astronomia, Universit\`a di Padova, Vicolo dell'Osservatorio, 3, I-35122, Padova, Italy\label{unipd}
         \and
              INAF Osservatorio Astronomico di Padova, vicolo dell'Osservatorio 5, I-35122 Padova, Italy\label{oapd}
         \and
              Dipartimento di Fisica e Astronomia, Universit\`a di Bologna,  via Gobetti 93/2,  40129 Bologna, Italy\label{unibo}
         \and
              INAF - Osservatorio di Astrofisica e Scienza dello Spazio di Bologna, Via Piero Gobetti, 93/3, 40129, Bologna, Italy\label{oabo}
         \and
              Department of Physics and Astronomy, University of Southampton, Highfield SO17 1BJ, UK\label{south}
         \and
              CEA, IRFU, DAp, AIM, Universit\'e Paris-Saclay, Universit\'e Paris Diderot, Sorbonne Paris Cit\'e, CNRS, F-91191 Gif-sur-Yvette, France\label{paris}
         \and
              Physics \& Astronomy Dept., University of Hawaii at Hilo., 200 W. Kawili St., Hilo, HI 96720, USA\label{uhawaii}
         \and
              Physics Dept., Carnegie Mellon University, Pittsburgh, PA 15213, USA\label{Carnegie}
         \and
              Department of Physics and Astronomy, Clemson University, Kinard Lab of Physics, Clemson, SC 29634, USA\label{clemson}
         \and
              School of Physics and Astronomy, Cardiff University, The Parade, Cardiff CF24 3AA, UK\label{cardiff}
         \and
              Center for Extragalactic Astronomy, Durham University, South Road, Durham DH1 3LE, UK\label{durham}
         \and
              European Southern Observatory, Alonso de Cordova 3107, Casilla 19, Santiago 19001, Chile\label{eso}
         \and
             Subaru Telescope, National Astronomical Observatory of Japan (NAOJ), 
             650 North A'ohoku place, Hilo, HI 96720, USA\label{subaru}
             }

   \date{Received ...; accepted ...}

 
  \abstract
   {}
   {We study the coevolution between the black hole accretion rate (BHAR) and the star formation rate (SFR) in different phases of galaxy life: main-sequence star-forming galaxies, quiescent galaxies, and starburst galaxies at different cosmic epochs.}
   {We exploited the unique combination of depth and area in the COSMOS field and took advantage of the X-ray data from the {\it Chandra} COSMOS-Legacy survey and the extensive multiwavelength ancillary data presented in the COSMOS2015 catalog, including in particular the UVista Ultra-deep observations.
   These large datasets allowed us to perform an X-ray stacking analysis and combine it with detected sources in a broad redshift interval ($0.1<z<3.5$) with unprecedented statistics for normal star-forming, quiescent, and starburst galaxies. 
   The X-ray luminosity was used to predict the black holeAR, and a similar stacking analysis on far-infrared {\it Herschel} maps was used to measure the corresponding obscured SFR for statistical samples of sources in different redshifts and stellar mass bins.
   }
   {We focus on the evolution of the average SFR-stellar mass (M$_*$) relation and compare it with the BHAR-M$_*$ relation. This extends previous works that pointed toward the existence of almost linear correlations in both cases. We find that the ratio between BHAR and SFR does not evolve with redshift, although it depends on stellar mass. For the star-forming populations, this dependence on M$_*$ has a logarithmic slope of $\sim0.6$ and for the starburst sample, the slope is $\sim0.4$. These slopes are both at odds with quiescent sources, where the dependence remains constant ($\log(\rm {BHAR}/{\rm SFR})\sim -3.4$).
   By studying the specific BHAR and specific SFR, we find signs of downsizing for M$_*$ and black hole mass (M$_{\rm BH}$) in galaxies in all evolutionary phases. The increase in black hole mass-doubling timescale was particularly fast for quiescents, whose super-massive black holes grew at very early times, while accretion in star-forming and starburst galaxies  continued until more recent times.}
   {Our results support the idea that the same physical processes feed and sustain star formation and black hole accretion in star-forming galaxies while the starburst phase plays a lesser role in driving the growth of the supermassive black holes, especially at high redshift. 
   Our integrated estimates of the M$_*$-M$_{\rm BH}$ relation at all redshifts are consistent with independent determinations of the local M$_*$-M$_{\rm BH}$ relation for samples of active galactic nuclei. This adds key evidence that the evolution in the BHAR/SFR is weak and its normalization is relatively lower than that of local dynamical M$_*$-M$_{\rm BH}$ relations.
   
   }


   \maketitle
%

\section{Introduction}
Observational studies and cosmological simulations have revealed a deep interconnection between galaxies and their central supermassive black hole (SMBH): they appear to coevolve and affect each other during their lives.
It has been reported several times that the SMBH mass correlates with a number of galaxy properties, including bulge mass, velocity dispersion, and S\'{e}rsic index. The relation with the velocity dispersion appears to the most fundamental relation so far \citep[e.g.,][]{2007ApJ...660..267B, 2016MNRAS.460.3119S, 2017MNRAS.466.4029S,2019MNRAS.485.1278S}.

Coincidentally, the cosmic star formation rate density (SFRD) and black hole accretion rate densities (BHARD) share a similar evolution: they reach a peak of activity at redshift $z\sim2$ and then decrease to the present epoch \citep{1998MNRAS.293L..49B, 2014MNRAS.439.2736D, 2014ARA&A..52..415M}. Additionally, semianalytic models and hydrodynamic simulations show that a self-regulating mechanism is required between the star formation and the black hole accretion in order to reproduce local scaling relations \citep[see][for a review]{2015ARA&A..53...51S}.
        
    The history of star formation in galaxies throughout the life of the Universe has been thoroughly constrained in recent years \citep[see][for a review]{2014ARA&A..52..415M}: the SFRD of the Universe increased by a factor of about 10 since $z\sim8$, reached a peak at around $z\sim2$, and declined by a factor of 10 since then. This decline is thought to depend on the decreasing availability of cold gas that is required to form stars \citep[e.g.,][]{2016MNRAS.458L..14F}. In addition, we know that most galaxies follow a sequence on the stellar mass ($M_*$) - star formation rate (SFR) plane with an almost linear slope. The so-called \emph{\textup{main-sequence}} of star-forming galaxies \citep{2004MNRAS.351.1151B, 2007ApJ...670..156D, 2007A&A...468...33E, 2007ApJ...660L..43N, 2014MNRAS.443...19R, 2017MNRAS.465.3390A} suggests that most of the cosmic star formation takes place through secular processes \citep{2011ApJ...739L..40R} because it has been observed at all redshifts and shows no evolution in its slope, but an increasing normalization with redshift \citep{2015A&A...581A..54T, 2015A&A...575A..74S, 2016ApJ...817..118T}: at earlier epochs, galaxies of a given stellar mass were forming more stars than in the local Universe.
    We also find that a number of galaxies lie substantially above and below the MS.
    Starburst galaxies are a small fraction of star-forming galaxies \citep[$\sim2\%$][]{2011ApJ...739L..40R}, and their SFRs and gas fractions are higher than those on the MS. 
    Quiescent galaxies, on the other hand, lie below the main-sequence and therefore have little star formation. They have been shown to evolve from $z\sim1.8,$ where they have significant amounts of dust and gas ($\sim5-10\%$) but their star formation efficiency is low, to the local universe, where they are gas poor \citep{2018NatAs...2..239G}.
    
    Just like star formation, black hole accretion depends on the availability of cold gas. Simulations show that gas flows into galaxies, where it cools to eventually fuel star formation and black hole accretion \citep{2010MNRAS.407.1529H}. This fueling by gas is expected to be even more pronounced in starburst galaxies, many of which are likely to undergo a major merging event \citep{2005Natur.433..604D, 2008ApJS..175..356H}: During the early stages of galaxy merging, gas can efficiently cool and lose angular momentum, eventually feeding central star formation and black hole growth. 
    The black hole accretion is initially obscured by thick layers of dust, which are then possibly removed by its increasing radiation and momentum feedback, revealing the quasar. Eventually, the gas is consumed, the quasar luminosity fades rapidly, and the star formation episode ceases. This leaves a "red and dead" elliptical galaxy with no or very little star formation or black hole accretion \citep{2004ApJ...600..580G, 2006ApJ...650...42L}.
    
    This picture might suggest that just like the majority of galaxies follow a main-sequence in the SFR-$M_*$ plane, a similar relation between the black hole accretion rate (BHAR) and the $M_*$ might exist: the more massive the galaxy, the higher the availability of inflowing gas  for star formation and black hole accretion, which would mean that they both should correlate with stellar mass. In addition, galaxies offset from the main-sequence (starbursts and quiescents), might have a BHAR that varies accordingly with the gas that is typically available in that phase \citep{2019ApJ...877L..38R}. 
    
    In order to search for these potential correlations, many authors have used the X-ray luminosity of galaxies as a proxy of BHAR: X-rays are very energetic photons that are created very close to the central SMBH, and other contaminants in the host galaxies at these wavelengths, for example, emission from stellar processes or binary systems, are usually less powerful and not dominant \citep[see the review by][]{2015A&ARv..23....1B}. Nevertheless, the first studies that traced the instantaneous BHAR with X-ray flux failed at finding any BHAR-M$_*$ relation \citep{2009ApJ...696..396S, 2010A&A...518L..26S, 2012MNRAS.419...95M, 2012A&A...545A..45R, 2015ApJ...806..187A}. A lack of a correlation between SFR and BHAR does not by itself necessarily imply a lack of physical connection. It might arise, for example, from different duty cycles and variabilities that characterize the two processes.  
    Episodes of star formation last for several Gyr, while the SMBH duty cycles are believed to be very short, with accretion episodes of about $10^5$~yr and variability timescales that range from minutes to months.
    
    In order to constrain more robust and reliable BHAR, 
    it is thus necessary to average their growth rate over a long time interval. A very promising technique to achieve this goal consists of stacking X-ray images. 
    Stacking allows us to perform studies on mass-complete samples by averaging the count rates of the X-ray images in the optical positions of the galaxies, thus increasing the signal-to-noise ratio, and allowing us to reach fluxes well below the single-source detection threshold of the observations. Moreover, because the BHAR is a stochastic event, stacking large samples of galaxies in a given volume is equivalent to averaging the growth rate of all galaxies.
    Previous works indeed searched for a relation between M$_*$ and average BHAR by stacking X-ray images \citep[e.g.,][]{2012ApJ...753L..30M, 2015ApJ...800L..10R, 2017ApJ...842...72Y}, and the probability distribution of specific X-ray luminosity with a maximum likelihood approach \citep{2012ApJ...746...90A, 2012MNRAS.427.3103B, 2018MNRAS.475.1887Y} and a Bayesian approach \citep{2018MNRAS.474.1225A}. All studies point toward a positive correlation between the BHAR and the M$_*$ for star-forming galaxies, very similar to the main-sequence of star-forming galaxies, with a slope close to unity, non-negligible redshift evolution, a positive slope for the BHAR-to-SFR ratio as a function of stellar mass, and indications of different behaviors for quiescent and starburst galaxies. 
    Thus far, no study has presented a complete analysis throughout all galaxy life phases by highlighting the evolution of the accretions and their ratio throughout cosmic time. This is what we present here. So far, only \citet{2019MNRAS.484.4360A} have shown that the fraction of active galactic nucleus (AGN) galaxies is higher below the main-sequence and in starbursts.
        
        We here characterize the evolution of the average BHAR for normal star-forming, quiescent, and starburst galaxies at $0.1<z<3.5$. This redshift interval encompasses the majority of the history of the Universe and contains two crucial epochs in its evolution: the peak of the star formation rate density and BHAR of the Universe at z$\sim$2, and their decline to the local Universe.
        We take advantage of the unique depth, area, and wavelength coverage of the COSMOS field, which allows us to select a mass-complete sample with large statistics out to very high redshifts for each galaxy phase. This is particularly important for starbursts, which are rare objects and require a large field in order to be found in good numbers for statistics. We apply the stacking technique to X-ray images from the \textit{Chandra} COSMOS-Legacy survey \citep{2016ApJ...819...62C} and combine the results with the actual X-ray detections in order to estimate the average X-ray luminosity and therefore average BHAR. We compare the evolution of the average BHAR with that of the average SFR. We show that these data confirm and extend previous claims about the evolution of the specific accretions and that the ratio of BHAR to SFR and M$_*$ are correlated. More specifically, in Section
 \ref{sec:data} we introduce the parent data sample we used for this work, from which we select the sample as described in Section \ref{sec:sample}. In Section \ref{sec:method} we describe the method we used to estimate the average X-ray luminosities, BHAR and SFR. In Sections~\ref{sec:L_X} and~\ref{sec:BH_SF} we describe our results and discuss them. In Section~\ref{sec:specifics} we describe the derivation of specific BHAR and specific SFR and discuss the respective results. In Section~\ref{sec:M_vs_M} we compare our M$_\text{BH}$-M$_*$ relation to observational and simulation relations. Finally, in Section~\ref{sec:conclusions} we enumerate the conclusions of this work.
   
Throughout this paper we use a \citet{2003PASP..115..763C} initial mass function (IMF) assuming a flat cosmology with $H_0=70$, $\Omega_\lambda=0.7$, $\Omega_0=0.3$.


\section{Data} \label{sec:data}
Our study focuses on the relation between average BHAR, SFR, and $M_*$ across a wide redshift range, from $z=0.1$ to $z=3.5$, and across different evolutionary stages of galaxies: (i) normal star-forming galaxies, which accrete gas secularly to slowly form new stars, and in which the bulk of the cosmic star formation took place; (ii) starburst galaxies, which are a small fraction of all galaxies at all epochs (\citealt[$\sim2\%$]{2011ApJ...739L..40R}; but see also \citealt{2017ApJ...849...45C}) 
and are going through a great burst of star formation that drives them significantly above the main-sequence galaxies on the SFR-M$_*$ plane;
and (iii) quiescent galaxies, which form stars at a very slow pace. 

Star-forming and quiescent galaxies can be easily selected at all redshifts from their emission in the optical/near-infrared (NIR) rest-frame bands, whereas starburst galaxies are heavily obscured in the optical and are therefore more easily identified by the thermal emission from the dust in the far-infrared (FIR). Therefore we selected our samples of star-forming and quiescent galaxies from the catalog by \citet{2016ApJS..224...24L}, which is NIR selected, while we used the FIR-selected catalog by \citet{2013MNRAS.432...23G} to identify starburst galaxies.
For the BHAR we used the catalog by \citet{2016ApJ...819...62C}, obtained from the \textit{Chandra} COSMOS-Legacy program, and complemented it by stacking on their X-ray images. 
We estimated the SFR for star-forming and quiescent galaxies by combining FIR stacking on \textit{Herschel} images and detections with UV luminosity.
We present in this section the catalogs we used for the sample selection and the data we used to estimate the average BHAR and SFR.

\subsection{Optical-NIR catalog}
We selected our sample of star-forming and quiescent galaxies in the COSMOS field \citep{2007ApJS..172....1S} from the COSMOS 2015 catalog \citep{2016ApJS..224...24L}, which uses photometry from UV (\textit{Galex}) to the mid-infrared (MIR, IRAC). This catalog constitutes the UltraVISTA DR2.

The multiwavelength catalog includes UV photometry in the far-UV and near-UV (NUV) bands with the \textit{GALEX} satellite \citep{2007ApJS..172..468Z}; UV/optical photometry in the u*-band from the Canada-France-Hawaii Telescope (CFHT/MegaCam); optical photometry from the COSMOS-20 survey, which is composed of 6 broad bands (B, V, g, r, i, z+), 12 medium bands (IA427, IA464, IA484, IA505, IA527, IA574, IA624, IA679, IA709, IA738, IA767, and IA827), and 2 narrow bands (NB711 and NB816), taken with the Subaru Suprime- Cam \citep{2007ApJS..172....9T, 2015PASJ...67..104T}; new and deeper z$^{++}$ and Y-band data, both taken with the Hyper-Suprime-Cam (HSC) on Subaru; H and $K_s$ NIR photometry obtained with WIRcam/CFHT \citep{2010ApJ...708..202M}, as well as deeper J, H, and $K_s$ imaging obtained with VIRCAM/VISTA UltraVISTA \citep{2012A&A...544A.156M} in the central 1.5~deg$^2$ of the COSMOS field (the coverage by
UltraVISTA is not homogeneous, with alternating "deep" and "ultradeep" stripes that reach depths of 24 and 24.7 in $K_S$-band, respectively); NIR data from IRAC, as part of the SPLASH COSMOS
, together with S-COSMOS \citep{2007ApJS..172...86S}; MIR and FIR data with \textit{Spitzer} IRAC and MIPS (Multi-band Imaging Photometer), from the \textit{Spitzer} Extended Mission Deep Survey and the \textit{Spitzer}-Candels survey \citep{2015ApJS..218...33A} data, among others; FIR from the \textit{Herschel} PACS and SPIRE instruments, taken as part of the PACS Evolutionary Probe (PEP) guaranteed-time key program \citep{2011A&A...532A..90L}, the largest field of the program, observed for about 200 h to a 3$\sigma$ depth at 160 $\mu m$ of 10.2 $mJy$ and at 100 $\mu m$ $\sim$5 $mJy$.

The point spread function (PSF) of the VISTA bands was estimated in each photometric band by modeling isolated known stars from the COSMOS ACS/HST catalog \citep{2007ApJS..172..196K, 2007ApJS..172..219L} with the PSFEX tool \citep{2013ascl.soft01001B}; 
The target PSF was chosen in order to minimize the applied convolutions, and it is the desired PSF of all bands after homogenization.
The required convolution kernel was calculated in each band by finding the kernel that minimizes the difference between the target PSF and the convolution product of this kernel with the current PSF. The images were then convolved with this kernel.

Then, a $\chi^2$ detection image \citep{1999AJ....117...68S}, produced by combining NIR images of UltraVISTA ($YJHK_S$) with the optical $z^{++}$-band data from Subaru, was used to identify objects; the photometry for these objects in the other bands was obtained by running SEXTRACTOR in dual-image mode. Fluxes were extracted from 2'' to 3'' diameter apertures on PSF-homogenized images in each band, except for a few cases: for \textit{GALEX}, the fluxes were measured using a PSF fitting method with the u*-band image used as a prior, and for the SPLASH IRAC imaging, the IRACLEAN tool \citep{2012ApJS..203...23H} was used to derive the photometry using the UltraVISTA $zYJHK_s$ $\chi^2$ image as a prior.
Photometry at 24 $\mu$m was obtained from the COSMOS MIPS-selected band-merged catalog \citep{2009ApJ...703..222L}. Far-IR photometry is provided by \textit{Herschel} PACS \citep[PEP guaranteed-time program,][]{2011A&A...532A..90L} and SPIRE \citep[HERMES consortium,][]{2012MNRAS.424.1614O} at 100, 160, 250, 350, and 500 $\mu$m.

The \citet{2016ApJS..224...24L} catalog provides secondary products as well, including photometric redshifts, stellar masses, and star formation rates. In particular, photometric redshifts were computed with \emph{Le Phare} \citep{2002MNRAS.329..355A, 2006A&A...457..841I} with the same method as was used in \citet{2013A&A...556A..55I}: the spectral energy distributions (SED) in 3 arcsec  apertures were fit to a set of 31 templates, including spiral and elliptical galaxies from \citet{2007ApJ...663...81P} and a set of 12 templates of young blue star-forming galaxies, produced using the models of \citet{2003MNRAS.344.1000B}. Photometric redshifts for objects that are detected in X-rays were instead taken from \citet{2016ApJ...817...34M}, who applied a set of templates that is more suitable for X-ray detected galaxies, in which the central black hole might significantly affect the UV/optical photometry (see section~\ref{ssec:Xray}). Photometric redshift precision was characterized by comparing it with spectroscopic samples from the COSMOS spectroscopic master catalog (M. Salvato et al., in preparation).

Stellar masses were derived using \emph{Le Phare}, with the same method as presented in \citet{2015A&A...579A...2I}: a \citet{2003PASP..115..763C} initial mass function was assumed, and the photometry was fit with a library of synthetic spectra generated using the stellar population synthesis model of \citet{2003MNRAS.344.1000B}, assuming both exponentially declining star formation histories (SFH) and delayed SFH ($\tau^{-2}te^{-t/\tau}$), assuming two different metallicities (solar and half-solar). Emission lines were added following the prescription in \citet{2009ApJ...690.1236I} together with two attenuation curves (the starburst curve of \citealt{2000ApJ...533..682C} and a curve with a slope $\lambda^{0.9}$ from Appendix A of \citealt{2013A&A...558A..67A}). The E(B - V) values were allowed to vary in a range between 0 and 0.7. The masses were then obtained as the median of the marginalized probability distribution function (PDF).

\subsection{IR data}\label{ssec:IRdata}
We selected the starburst galaxies sample for this work from a FIR-selected catalog. In particular, we used the \citet{2013MNRAS.432...23G} catalog, which is selected from PACS/\textit{Herschel} PEP observations in the COSMOS field. This catalog was also matched with the deep 24 \textmu m imaging of \citet{2009ApJ...703..222L}, with the HerMES extragalactic survey \citep{2012MNRAS.424.1614O} observed with SPIRE at 250, 350, and 500~\textmu m in the same fields that were covered by PEP and with the IRAC-based catalog of \citet{2010ApJ...709..644I}, including optical and NIR photometry and photometric redshifts.

\citet{2013MNRAS.432...23G}) used as reference the blind catalogs at 100 and 160 \textmu m from \citet{2010A&A...518L..30B,2011A&A...532A..49B}, selected down to the $3\sigma$ level, which in COSMOS contain 5355 and 5105 sources at 100 and 160 \textmu m, respectively. Then they associated their sources with the ancillary catalogs by means of a multiband likelihood ratio technique \citep{1992MNRAS.259..413S,2001Ap&SS.276..957C}, starting from the longest available wavelength (160 \textmu m, PACS) and progressively matching 100 \textmu m (PACS) and 24 \textmu m (MIPS). Their final catalog consists of 4110 and 4118 sources at 100 and 160 \textmu m, respectively. Spectroscopic or photometric redshifts are available for 3817 and 3849 of their sources, respectively, which they used for the SED fitting in order to estimate the FIR luminosity function.

Stellar masses were obtained by fitting the broadband SED with a modified version of MAGPHYS \citep*{2008MNRAS.388.1595D}, which  simultaneously fits the broadband UV-to-FIR observed SED of each object and ensures an energy balance between the absorbed UV light and the light that is reemitted in the FIR regime. The redshift for each object was fixed to the spectroscopic redshift when available or else to the photometric redshift; then the SED was fit to a best-fit model that we selected from a library built by combining different SFH, metallicities, and dust contents. Each SFH is the combination of an exponentially declining SFR model, to which random bursts of star formation are superimposed \citep[see][]{2008MNRAS.388.1595D, 2010MNRAS.403.1894D}. The emission of a possible AGN component was taken into account using a modified version of the MAGPHYS code \citep[SED3FIT,][]{2013A&A...551A.100B}, which adds a torus component to the modeled SED emission by combining the \citet{2008MNRAS.388.1595D} original code with the \citet*{2006MNRAS.366..767F} AGN torus library \citep[see also][]{2012MNRAS.426..120F}.

In order to estimate the SFR of their sample, \citet{2013MNRAS.432...23G} calculated an SED using all the available multiwavelength data by performing a $\chi^2$ fit using the \emph{Le Phare} code \citep{2002MNRAS.329..355A, 2006A&A...457..841I} with the semiempirical template library of \citet{2007ApJ...663...81P}, which is representative of different classes of IR galaxies and AGN. They also added some modified templates in the FIR to better reproduce the observed {Herschel} data \citep[see][]{2010A&A...518L..27G}, and three starburst templates from \citet{2009ApJ...692..556R}.
Then they integrated the best-fitting SED of each source over $8\leq \lambda_{rest}\leq1000$~\textmu m to derive the total IR luminosities (L$_{\rm IR}$ = L[8--1000 \textmu m]) in 11 redshift bins (0.0--0.3, 0.3--0.45, 0.45--.6, 0.6--0.8, 0.8--1.0, 1.0--1.2, 1.2--1.7, 1.7--2.0, 2.0--2.5, 2.5--3.0, and 3.0--4.2). 
Finally, they estimated the SFR$_{\rm IR}$ for all these sources from the total IR luminosity after subtracting the AGN contribution and using the \citet{1998ARA&A..36..189K} relation \citep[see also][for more details]{2015MNRAS.451.3419G}.

\subsection{X-ray data} \label{ssec:Xray}
We used X-ray data from the \textit{Chandra} COSMOS Legacy Survey \citep[COSMOS-Legacy,][]{2016ApJ...819...62C}, a 4.6 Ms \textit{Chandra} program that combines new observations obtained during \textit{Chandra} Cycle 14 with the previous C-COSMOS Survey, allowing the X-ray data to uniformly cover the whole 2.2 deg$^2$ of the COSMOS field. The limiting fluxes are  $2.2 \times 10^{-16}, 1.5 \times 10^{-15}$, and $8.9 \times 10^{-16} \um{erg}\ump{cm}{-2} \ump{s}{-1}$ in the 0.5--2, 2--10, and 0.5--10~keV bands, respectively.

The optical counterparts to the X-ray COSMOS-Legacy sources are available in \citet{2016ApJ...817...34M}. This was obtained by using the maximum likelihood ratio technique \citep[e.g.,][]{1992MNRAS.259..413S, 2005A&A...432...69B, 2012ApJS..201...30C} by matching X-ray sources to three separate bands: i-band data from \citet{2009ApJ...690.1236I}, $K_s$-band data using UltraVISTA DR2, and IRAC 3.6 micron using either SPLASH or \citet{2007ApJS..172...86S}
sources from the \citet{2016ApJS..224...24L} catalog in the UltraVISTA field.
Optical counterparts were cross-correlated with the master spectroscopic catalog (Salvato et al. in prep), which contains spectroscopic redshifts from numerous observing campaigns and instruments. For the sources for which no spectroscopic redshift is available, a photometric redshift was provided that we obtained by following the same procedure as in \citet{2011ApJ...742...61S}. 

   \begin{figure*}
   \centering
   \includegraphics[trim={2cm 6cm 1cm 5.5cm}, clip, width=0.9\paperwidth]{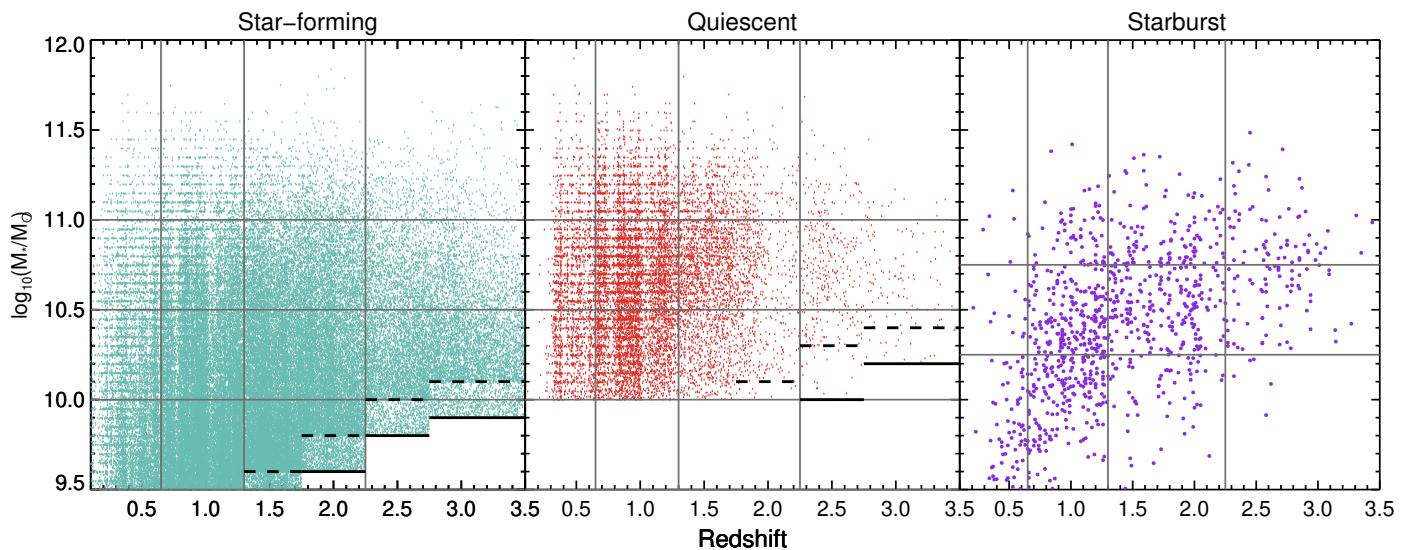}
    \caption{ Stellar mass of our sample of galaxies shown as a function of redshift. Star-forming and quiescent galaxies are selected from \citet{2016ApJS..224...24L} (green and red data points in the left and central panel, respectively), while the starbursts are selected from \citet{2013MNRAS.432...23G} (violet data points in the right panel).
                        The solid and dashed black lines are the mass completeness thresholds are for UVista deep and ultradeep stripes from \citet{2016ApJS..224...24L} for star-forming and quiescent galaxies. For starburst galaxies we considered as mass incomplete the low-mass bins (9.50<$\log_{10}($M$_\sun$/M$_*)$<10.25) at $z>1.30$.
                The solid gray horizontal and vertical lines show the limits of our mass and redshift bins.
                }
         \label{fig:M_vs_z}
   \end{figure*}


\section{Sample selection} \label{sec:sample}
We selected a mass-complete sample from \citet{2016ApJS..224...24L} by limiting our selection to the area that is covered by the UltraVISTA-DR2 observations (1.5 deg$^2$) in the redshift interval $0.1<z<3.5$. In order to separate star-forming from quiescent galaxies, we used the classification from \citet{2016ApJS..224...24L}, which  is based on the rest-frame NUV$ - r / r - J$ color-color diagram as in \citet{2013A&A...556A..55I}. It allows separating dust-obscured galaxies from older stellar populations. In this diagram, galaxies with colors NUV$ - r>3 (r-J)+1$ and NUV$ - r > 3.1$ are classified as quiescent.

We selected starburst galaxies from the \citet{2013MNRAS.432...23G} catalog in the same redshift interval down to M$_*\sim 10^{9.5} M_\sun$. Again, we chose this catalog because it is FIR selected, that is, selected based on the specific star formation rate (at least to a first approximation), and includes SFR$_{\rm IR}$ estimates, which allowed us to robustly estimate the SFR of highly star-forming galaxies that are heavily obscured. 
Based on Fig. 15 of \citet{2013MNRAS.432...23G}, above $z\sim1.3$ we probably miss the low-SFR starburst sources, therefore we decided to show these data points as upper limits.
 The starburst galaxies were selected to have an SFR that is at least four times higher than the SFR of a typical main-sequence galaxy with a similar stellar mass and redshift, which is our definition of starburst galaxy.
For the purpose of their selection, we considered the parameterization of the main-sequence from \citet{2015A&A...575A..74S} and obtained the relation
\begin{multline}  \label{eq:SFR_thresh}
\textrm{SFR}_{IR}^{SB} \geq 4 \times \textrm{SFR}_{MS}(z_{ave},M_*)=\\
=4 \times \left[m-m_0+a_0r-a_1[\max(0,m-m_1-a_2r)]^2 \right],
\end{multline}
where SFR$_{IR}^{SB}$ is the star formation rate of a starburst galaxy obtained from its IR luminosity, $r=\log_{10}(1+z)$, $m=\log_{10}(M_*/10^9M_\cdot)$, and the parameters $m_0=0.5$, $m_1=0.36$, $a_0=1.5,$ and $a_1=0.3$.
We used SFR$_{IR}^{SB}$, M$_*$ , and $z$ from the \citet{2013MNRAS.432...23G} catalog.
We matched the starburst sample with the catalog of star-forming and passive galaxies that was previously color-color selected from \citet{2016ApJS..224...24L}, using a 3.5'' association radius in order to exclude the starburst galaxies from our star-forming and quiescent selection. For a distance greater than 2'' we only accepted matches that have a $\Delta z = |\, z_{\rm Gruppioni} - z_{\rm Laigle} \,|<0.3$, allowing us to include seven more starburst galaxies in our sample.
We compared the stellar masses of the starburst sample (derived from MAGPHYS) with their stellar masses from \citet{2016ApJS..224...24L} (derived from \emph{Le Phare}) in each of the redshift bins and found that they lie along the 1:1 relation with increasing scatter with redshift (RMS ranging from 0.18 dex at low z to 0.35 dex at high-z), but with no offset (the median of the difference of the masses is about $10^{-2}$). From the cross-match we find that 965 color-selected star-forming galaxies are IR-classified as starbursts (this is expected because the NUV$ - r / r - J$ diagram does not have a starburst area) and 28 color-selected quiescents are IR starbursts. We moved 1494 galaxies that are $24\mu m$ MIPS/\textit{Spitzer} detected from our quiescent selection to the star-forming galaxy sample because the IR emission indicates that a certain amount of optically hidden star formation is ongoing. For the X-ray detected star-forming and quiescent galaxies we used the redshift included in \citet{2016ApJ...817...34M}. We have a final number of 83,904 star-forming, 12,839 quiescent, and 1,003 starburst galaxies.

We further divided the sample into four redshift bins: 0.1<z<0.65, 0.65<z<1.3, 1.3<z<2.25, and 2.25<z<3.5.

Every redshift bin was chosen with the purpose of including a sufficient number of galaxies (see Table~\ref{table:n_gal}), and then the galaxy sample of each redshift bin was divided into mass bins. Our final sample is shown in Fig.~\ref{fig:M_vs_z} together with the mass-completeness thresholds from \citet{2016ApJS..224...24L} for star-forming and quiescent galaxies and the limits of our stellar mass and redshift bins. 

\section{Method} \label{sec:method}
The aim of this paper is to study the average BHAR for galaxies with different star formation activities, from passive galaxies to starbursts, at different cosmic epochs and at different stellar masses. Because AGN activity is a stochastic event, as the AGN duty cycle is short compared to the star formation episodes in galaxies, it is unlikely that a galaxy has been observed at the peak of its black hole accretion. This means that if we limited the analysis to bright X-ray galaxies, we would obtain a biased view of the average AGN activity in
galaxies. For this reason, we complement X-ray individual detections with the stacking analysis on \textit{Chandra} X-ray images for nondetected galaxies. 

\subsection{X-ray stacking analysis} \label{sec:stacking}
In order to estimate the average X-ray luminosity for each mass and redshift subsample, we followed the same method as in \citet{2015ApJ...800L..10R}. We used individual X-ray detections from the COSMOS-Legacy catalog \citep{2016ApJ...819...62C} when possible, taking advantage of the match between optical and X-ray sources performed by \citet{2016ApJ...817...34M} (see section \ref{ssec:Xray}). For the non-X-ray detected sources we performed stacking on the same images using the CSTACK tool v4.32\footnote{\url{http://cstack.ucsd.edu/cstack} or \url{http://lambic.astrosen.unam.mx/cstack/} developed by Takamitsu Miyaji.}  \citep{2008HEAD...10.0401M}.

We propagated the probability distributions of count rates with Monte Carlo simulations. We focused on the 2-7~keV band and took advantage of the CSTACK bootstrap output. We simulated it $10^6$~times by interpolating the inverted cumulative distribution function of the bootstrap probability distribution.
For X-ray detections, we instead assumed a Gaussian probability distribution and a $\sigma$ equal to the error on the count rate from the catalog when available. Alternatively, we set the error on the detection to $\sigma=0.25 \times CR(2-7\um{keV}).$  Count-rate distributions from the stacking and detections were converted into standard flux in the $2-10\um{keV}$ band through the WebPIMMS\footnote{\url{http://cxc.harvard.edu/toolkit/pimms.jsp}} conversion factor obtained for \textit{Chandra} cycle 14
by assuming a power-law spectrum for the AGN with photon index $\Gamma$= 1.8 and galactic hydrogen absorption \citep[N$_H=2.6\times 10^{20} \ump{cm}{-2}$,][]{2005A&A...440..775K}.
This value of the photon index $\Gamma$ is compatible with the cosmic X-ray background according to \citet{2019ApJ...871..240A} for cutoff energies E$_c<100\um{keV}$.
We converted fluxes into rest-frame with a K-correction with the shape of $K_{corr}=(1+z)^{\Gamma-2}$ , where $z$ is the average redshift of each subsample and $\Gamma$ is the photon index just introduced.

Because we search for a global average X-ray measurement by combining detections and non-detections in each redshift and mass bin, we applied the following equation to all the fluxes in the probability distribution of each detection and the stacking:
\begin{equation}  \label{eq:F_ave}
\centering
F_\text{ave}=\frac{\sum_{j=1}^{n_{\text{detected}}}F_{j,\text{detected}}+n_{\text{stacked}}\cdot F_{\text{stacked}}}{n_{\text{detected}}+n_{\text{stacked}}}
,\end{equation}
where in each redshift and mass bin, $n_{\text{detected}}$ is the number of detections, $F_{j,detected}$ is the flux of each detection, $n_{\text{stacked}}$ is the number of undetected objects that were therefore stacked, and $F_{stacked}$ is the total flux from the stacked objects. This returns a probability distribution of rest-frame flux for each redshift and mass bin. We show in Table \ref{table:n_gal} the number of stacked and detected galaxies.

   \begin{figure*}
   \centering
   \includegraphics[trim={1cm 2cm 2.5cm 3cm}, clip,width=0.9\paperwidth]{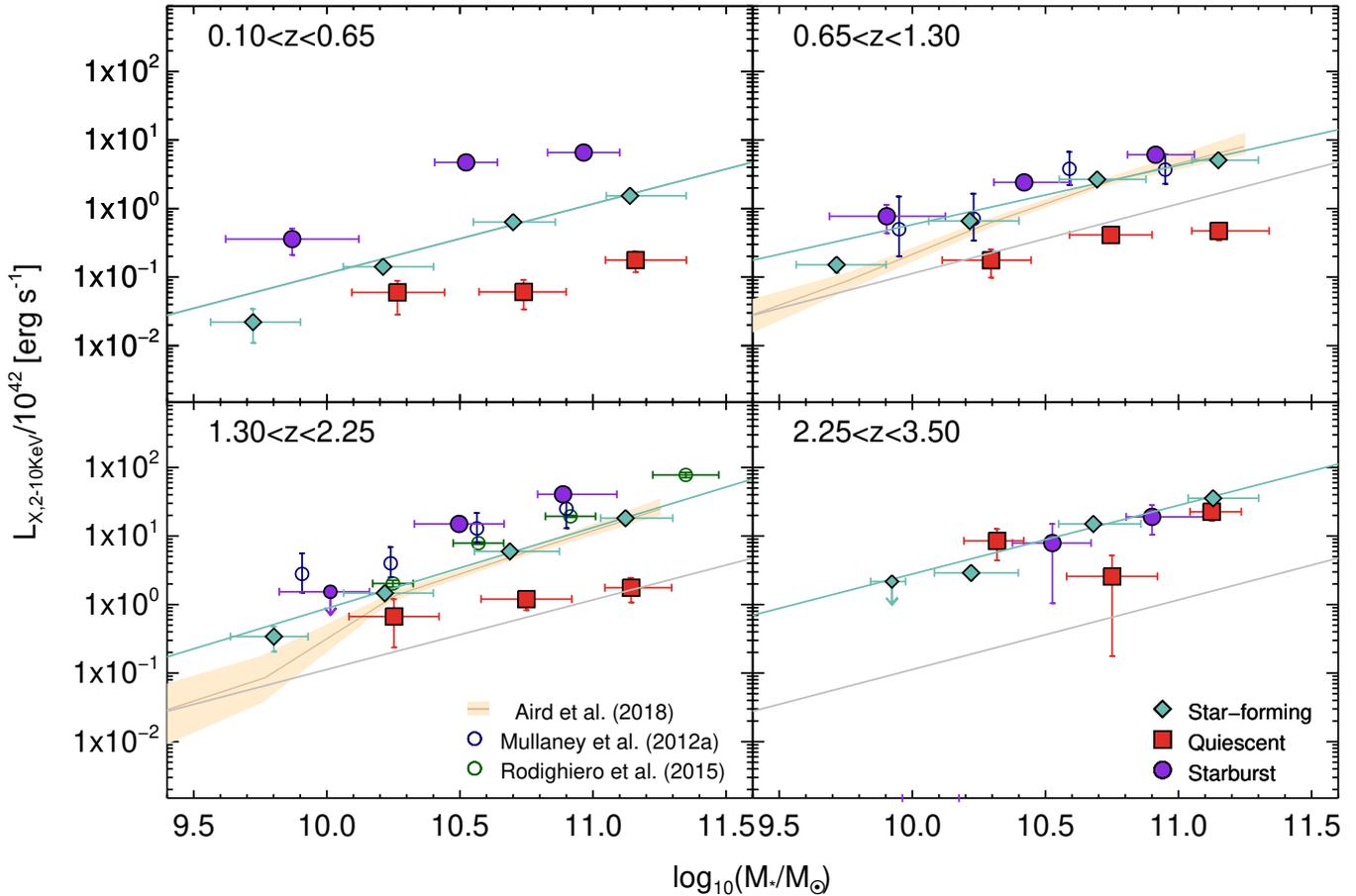}
      \caption{X-ray luminosity in the 2-10~keV band for our sample as a function of stellar mass. Each panel represents one of the four redshift bins that we considered. The light green diamonds represent the normal star-forming sample, the red squares represent the quiescent sample, and the violet circles represent the starburst sample. The data points represent the median value of the stellar mass and the X-ray luminosity. Vertical error bars represent the 5th and 95th percentile (90\% confidence interval) of the distributions, while horizontal error bars correspond to $1~\sigma$. We show an upper limit (the 95th percentile) for the data points whose X-ray luminosity 5th percentile is compatible with zero and for starburst galaxies at high redshift and low mass, where our sample is incomplete.
      The light green continuous line is the best fit for the star-forming galaxy data points at each redshift, while the gray line shows the best fit in the lowest redshift bin, shown for comparison purposes.  
      We also report data points from \citet{2012ApJ...753L..30M} at an average redshift of $z=1,2$ (open dark blue circles, \citet{2015ApJ...800L..10R} at an average redshift of $z=2$ (open green circles) and \citet{2018MNRAS.474.1225A} at $z=1,2$ (solid brown curve; the yellow area shows the $1\sigma$ confidence interval). The data points from \citet{2015ApJ...800L..10R} and \citet{2018MNRAS.474.1225A} are scaled to the same $k$-correction as we adopted here.
              }
         \label{fig:L_X}
   \end{figure*}
%
%
\begin{table*}
\caption{Number of X-ray (2-7~keV band) stacked and detected galaxies per redshift bin and galaxy type. 
}             
\label{table:n_gal}      
\centering                          
\begin{tabular}{c | c c | c c |c c }        
\hline\hline                 
              & Star-forming &         & Quiescent  &         & Starburst  &        \\
Redshift bins & detections   & Stacked & detections & Stacked & detections & Stacked\\
\hline                                          
$0.10<z<0.65$ & 249          & 8,211   & 45         & 1,968   & 12         & 66    \\
$0.65<z<1.30$ & 743          & 28,037  & 116        & 7,055   & 22         & 366    \\
$1.30<z<2.25$ & 764          & 31,012  & 42         & 2,835   & 48         & 280    \\
$2.25<z<3.50$ & 337          & 11,615  & 16         & 443     & 9          & 115     \\
\hline                                   
\end{tabular}
\end{table*}
    
%
%
\begin{table}
\caption{X-ray luminosity fits to the equation $\log_{10}L_X=a\log_{10}{\rm M}_*+ b$ for the star-forming galaxies of our sample. Errors are $1\sigma$ uncertainty estimates of parameters}             
\label{table:lx_fitpar}      
\centering                          
\begin{tabular}{c c c }        
\hline\hline                 
Redshift bin  & $a$               & $b$       \\    
\hline                        
$0.10<z<0.65$ & $ 1.02 \pm 0.03 $ & $-11.1 \pm 0.3$ \\
$0.65<z<1.30$ & $ 0.87 \pm 0.02 $ & $ -8.9 \pm 0.2$ \\
$1.30<z<2.25$ & $ 1.18 \pm 0.03 $ & $-11.8 \pm 0.3$ \\
$2.25<z<3.50$ & $ 1.00 \pm 0.05 $ & $ -9.6 \pm 0.5$ \\
\hline                                   
\end{tabular}

\end{table}

\subsection{X-ray luminosity and black hole accretion rate estimate}
We converted the distribution of average rest-frame fluxes F$_\text{ave}$  
(Equation~\ref{eq:F_ave}) to average luminosities L$_\text{ave}$ 
 using the corresponding luminosity distance at the average redshift of the sources included in the bin. We estimated and subtracted the contribution from the stars (young and old stellar populations) in the 2-10~keV range as in \citet{2016ApJ...825....7L} using their best-fit values,
\begin{equation}  \label{eq:XSFR}
\centering
\text{L}_{2-10\um{keV}}(M_*,z,\text{SFR})[\um{erg}\ump{s}{-1}]=\alpha_0(1+z)^\gamma M_*+\beta_0(1+z)^\delta \text{SFR}
,\end{equation}
with $\log\alpha_0=29.37$, $\log\beta_0=39.28$, $\gamma= 2.03,$ and $\delta=1.31$. This correction has two terms that take into account the X-ray emission from the young stellar populations, that is, high-mass X-ray binaries, whose emission is proportional to the SFR of the galaxy (see Sec.~\ref{ssec:SFR}), and from the old stellar populations, that is, low-mass X-ray binaries, whose emission is proportional to the stellar mass of the galaxy. This is particularly important in quiescent galaxies. We confirmed the effect of this correction by comparing our results with the corrections  from \citet{2018ApJ...865...43F} and \citet{2017MNRAS.465.3390A}. The correction from \citet{2017MNRAS.465.3390A} gives no apparent difference to our results, while the correction from \citeauthor{2018ApJ...865...43F}, which is to be considered an upper limit because it includes some residual AGN emission, has some effect at the lowest redshift that causes some data points to become compatible with zero and a lower normalization by $\sim0.2$~dex in star-forming and quiescent galaxies. The overall results are not affected by it, however.



In order to correct for the average internal absorption from the galaxy (N$_H$), we estimated the hardness ratios (HR) of our sample. We define the HR as
\begin{equation}  \label{eq:HR}
\centering
\text{HR}=\frac{H-S}{S+H},
\end{equation}
where $H$ and $S$ are the median of the distributions of count rates in the 2-7~keV and 0.5-2~keV band, respectively. 
We compared the estimated HRs of our data with those obtained from models in PyXspec. The model we used is PHABS*(CABS*ZPHABS*PO+APEC), which is composed of a redshifted and absorbed power-law emission with slope $\Gamma=1.8$ and hot gas emission at $kT=1$~keV. The model also includes galactic absorption. The power-law spectrum was normalized because the fluxes we measured and the APEC warm gas component were normalized to equal the emission expected from the SFR of the galaxies \citep[][Eq.~1]{2012MNRAS.426.1870M}. Finally, we convolved the model with an auxiliary response file (ARF) from \textit{Chandra} ACIS-I cycle 14 in order to obtain the $H$ and $S$ count rates. All models give two N$_H$ values at which the HRs are compatible with our data. The only exception are star-forming galaxies in the lowest redshift bin, where the modeled spectra are too soft, probably because the 1~keV gas emission is overestimated. In the remaining cases, we find a possible solution at N$_H = 10^{22.0-23.2}\ump{cm}{-2}$ and another at higher N$_H$ values N$_H = 10^{23.5-24.8}\ump{cm}{-2}$ , where all the power-law emission has been absorbed and only the hot gas is visible. The two N$_H$ solutions increase with redshift.
We decided to use the solution with a lower N$_H$ because we cannot observe galaxies with Compton-thick levels of obscuration. Even though we performed stacking, the number of photons from heavily obscured AGN in the energy range of \textit{Chandra} is very low and does not dominate our population. Furthermore, our NIR selection may already exclude dust-obscured galaxies, which are thought to be heavily absorbed in the X-rays \citep{2008ApJ...672...94F, 2016A&A...592A.109C, 2019AJ....157..233R}. A last consideration is that even though many works reported an intrinsic fraction of highly obscured AGN or Compton-thick AGN of about $\sim30-50\%$ at various redshifts \citep{2014ApJ...786..104U, 2014MNRAS.445.3557V,2015ApJ...815L..13R,2019ApJ...871..240A}, it still is not possible to detect these galaxies in the \textit{Chandra} X-ray energies.
The HRs of our sample do not show a trend with mass, therefore we decided to use the same obscuration correction for all mass bins of a given galaxy type and redshift.
We estimated the correction factors using WebPIMMS, and it extends from absorbed to unabsorbed flux in the 2-10~keV band at the average redshift of the bin and with the N$_H$ found from the HR analysis for each galaxy type. These corrections are approximately $5-10\%$.

Finally, we considered the median value of the X-ray luminosity distribution as the representative luminosity of each bin and the 5th and 95th percentiles as the lower and upper limits to the uncertainty associated with the X-ray luminosity. We considered values compatible with zero when the 5th percentile assumed a negative value, and in these cases, we show them as upper limits.

We transformed the X-ray binaries and obscuration-corrected X-ray luminosity into a BHAR as in \citet{2008MNRAS.388.1011M}, \citet{2012ApJ...753L..30M}, \citet{2014MNRAS.439.2736D}, \citet{2015ApJ...800L..10R}, and \citet{2018ApJ...857...64B} with the relation
\begin{equation}  \label{eq:BHAR}
\centering
\text{BHAR}(\text{M}_*,z)=\frac{(1-\epsilon)\times\text{L}_{\text{bol}}(\text{M}_*,z)}{\epsilon c^2},
\end{equation}
where L$_{\text{bol}}$ is the AGN bolometric luminosity obtained using a luminosity-dependent bolometric correction consisting of (i) the bolometric correction used in \citet{2018MNRAS.475.1887Y}, which is a modified version of the bolometric correction from \citet{2012MNRAS.425..623L}, down to $\text{L}_\text{X}=10^{42.4} \um{erg}\ump{s}{-1}$ , and (ii) $k_{bol}=16$ for lower luminosities as in \citet{2017ApJ...842..131S}.
$c$ is the speed of light in vacuum, and $\epsilon$ is the efficiency by which mass is converted into radiated energy in the accretion process. Here we assumed $\epsilon=0.1$ \citep[e.g.,][]{2004MNRAS.351..169M, 2012ApJ...753L..30M, 2015ApJ...800L..10R, 2018ApJ...857...64B}, or that roughly 10\% of the accreted rest-mass is converted into radiant energy, regardless of M$_\text{BH}$.

\subsection{Star formation rates: FIR stacking and UV SED fitting} \label{ssec:SFR}
 For star-forming and quiescent galaxies, we estimated the SFR as the sum of the IR and far-UV-observed contributions (SFR$_{\rm IR + UV}$), while for starburst galaxies, we only considered the SFR$_{\rm IR}$. The SFR$_{\rm IR}$ was derived through the empirical calibration of \citet[][Eq.~4]{1998ARA&A..36..189K} from the total IR luminosity (L$_{\rm IR}$). The SFR$_{\rm UV}$ (not corrected for extinction) was inferred, following the prescription of \citet[][Eq.1]{1998ARA&A..36..189K}, from the rest-frame luminosity at 1600~\AA~ ($L_{1600\AA}$). The two SFRs were then combined as discussed by \citet{2013ApJ...762..125N}, and they were finally converted into the \citet{2003PASP..115..763C} IMF as in \citet{2008A&A...482...21C} by dividing by a factor of 1.7,
\begin{multline}   \label{eq:SFR_tot}
\text{SFR}_{\rm IR + UV}[{\rm M}_\odot \; yr^{-1}]= \text{SFR}_{\rm UV} + \text{SFR}_{\rm IR} = \\
=(2.86  \;{\rm L}_{1600\AA}+1.7 \; {\rm L}_{\rm IR})\times \frac{10^{-10}}{1.7} \; [{\rm L}_\odot].
\end{multline}
 
 In order to obtain the SFR$_{\rm IR + UV}$ for star-forming and quiescent galaxies, we performed bootstrapping, in which we selected a subsample of galaxies in a M$_*$ and $z$ bin at each iteration and estimated their L$_{\rm IR}$ and L$_{1600\AA}$. We then combined the median of each luminosity through Eq.~\ref{eq:SFR_tot} and thus derived the SFR$_{\rm IR + UV}$ distribution.  L$_{\rm IR}$ was obtained by stacking $160~\mu m$ PACS/\textit{Herschel} maps as in \citet{2014MNRAS.443...19R}. In our procedure we accounted for detections and nondetections to obtain the final median stacked fluxes that we then converted into bolometric luminosities by adopting an average $k$-correction \citep{2001ApJ...556..562C}. Because the $L_{1600\AA}$ luminosity is not included in \citet{2016ApJS..224...24L}, we derived it in the following way. For each source in the sample, we reconstructed the galaxy SED using {\it Hyperzmass}, which is a modified version of the {\it Hyperz} software \citep{2000A&A...363..476B, 2010A&A...524A..76B} and is suitable for computing the stellar mass when the photometric redshift is known. We adopted the \citet{2003MNRAS.344.1000B} stellar population models, with exponentially declining SFH, that is, SFR~$\propto e^{t/\tau}$), with $\tau$=0.1, 0.3, 1, 2, 3, 5, 10, 15, 30,  $\infty$=constant SFR). The $L_{1600\AA}$ were extracted directly from the {\it Hyperzmass} best-fit templates. 
The SFRs for quiescent and star-forming galaxies are shown in the upper left and central panels of Fig.~\ref{fig:SF_BH_all} as a function of stellar mass. They are color-coded based on redshift.
 
 For starburst galaxies we used the SFR$_{\rm IR}$ from \citet{2013MNRAS.432...23G} (see Section \ref{ssec:IRdata}) after  verifying that their SFR$_{\rm UV}$ is negligible. 
 Similarly to star-forming and quiescent galaxies, we performed bootstrapping for every M$_*$ and $z$ bin, where in each loop we derived a median SFR$_{\rm IR}$ and therefore an SFR$_{\rm IR}$ distribution.
 The SFRs for starburst galaxies are shown in the top right panel of Fig.~\ref{fig:SF_BH_all}.

\section{Comparing the SMBH X-ray emission at the different galaxy life stages throughout cosmic time}\label{sec:L_X}
In Figure \ref{fig:L_X} we show the X-ray 2-10~keV luminosity (L$_\text{X}$) from Equation \ref{eq:XSFR} as a function of stellar mass for our sample, divided into different redshift bins. For each bin we compared the X-ray luminosity for the three types of galaxies: star-forming, quiescent, and starbursts. Our data points are centered on the median X-ray luminosity and stellar mass. Vertical error bars are the 90\% confidence range. We show a log-log linear fit for star-forming galaxy data points. The fits were performed ignoring the upper limit data points.
   
\begin{figure*}
   \centering
   \includegraphics[trim={3.2cm 3.7cm 2cm 3.6cm}, clip,width=\linewidth]{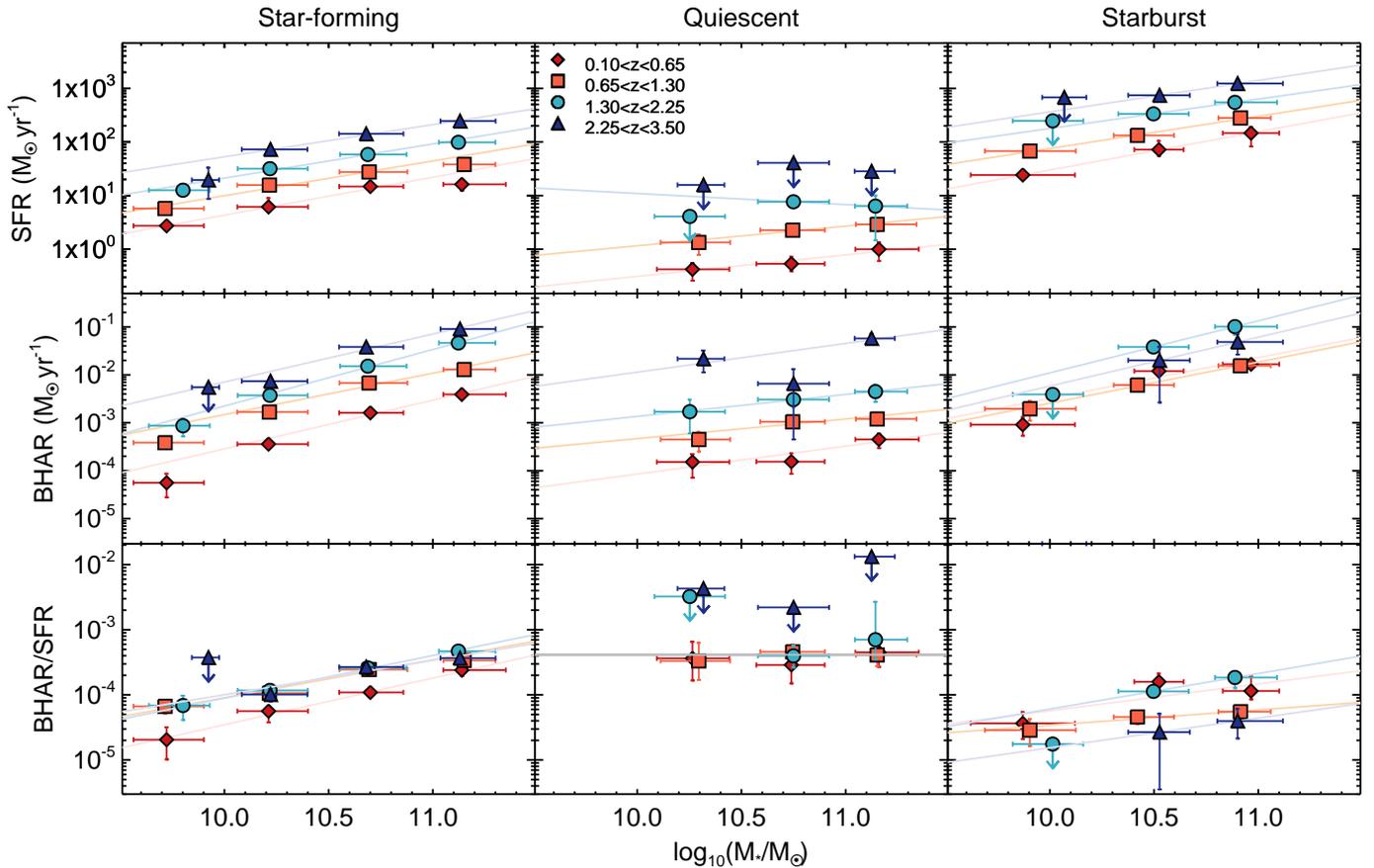}
      \caption{Comparison between the average SFR and the average BHAR for the three samples: normal star-forming galaxies (left), quiescent galaxies (center), and starburst galaxies (right).
      In all plots data points are color- and shape-coded according to the redshift interval. The data points are the median value, and vertical errors represent the 5th and 95th percentile. We show an upper limit (the 95th percentile) for the data points whose SFR or BHAR 5th percentile is compatible with zero and for starburst galaxies at high redshift and low mass, where our sample is inclomplete in mass.
      The error bars shown on the stellar masses correspond to $1~\sigma$ of the distribution. The solid light colored lines are the best fits and are color-coded according to each redshift bin (see Table~\ref{table:all_fitpar}). 
      Top panel: M$_*$-SFR relation. 
      Middle panel: M$_*$-BHAR relation.
      Bottom panel: BHAR-to-SFR ratio as a function of stellar mass. The dotted gray line for quiescent galaxies is at a constant value of BHAR/SFR$=3.8\times 10^{-4}$.
              }
         \label{fig:SF_BH_all}
   \end{figure*}

We find a robust relation between the average L$_\text{X}$ and M$_*$ in star-forming galaxies: this indicates that black holes in more massive galaxies grow faster than black holes in the less massive galaxies. This might in part be sustained by the higher fraction of type I AGN at high stellar masses, as seen by \citet{2019ApJ...872..168S} for \textit{Chandra} COSMOS Legacy detected sources. 
Moreover, this relation increases in normalization by about 1.5 dex but maintains an almost constant slope of about $\sim1.00$ toward higher redshift, which indicates that black holes grew faster at earlier epochs than they do today.  This is shown in Table \ref{table:lx_fitpar}, which lists the best linear fit to our data ($\log\text{L}_X=a\log\text{M}_*+b$).
This relation looks very similar to the main-sequence of star-forming galaxies, which is the relation between SFR and M$_*$, as we show more clearly in Figure \ref{fig:SF_BH_all}. The quiescent $24\mu m$ MIPS/\textit{Spitzer} detected sources are a small percentage of the overall star-forming sample ($\sim2$\%), but they are mostly concentrated at the high masses and low redshifts, where they reach a percentage as high as 86\% at M$_*>10^{11}$M${_\odot}$ in the lowest redshift. The addition of these sources had very little effect on the L$_\text{X}$ of the higher mass bins, but all bins show a systematical increase in L$_\text{X}$  of~$\lesssim 0.1$~dex.

Starburst galaxies show average L$_\text{X}$ with a similar dependence on stellar mass as star-forming galaxies and a mild dependence on redshift: as a result, the X-ray luminosities of starbursts are $\sim0.4-1.1$~dex higher than those of star-forming galaxies at $z\simeq0.4$, but the luminosities at $z\simeq2-3$ are compatible.
If these galaxies undergo a major merger event, as is commonly believed, this suggests that while a higher availability of cold gas allows the SFR to increase considerably with respect to a galaxy in a secular evolution phase, this gas availability is not able to accrete onto the black hole at a pace higher than a certain threshold that does not vary as much as the star-forming population ($\sim0.5$~dex) in the redshift range we covered.

Finally, quiescent galaxies have X-ray luminosities that tend to be lower than those of star-forming galaxies. They vary with increasing mass from $\sim0.5$~dex to $\sim1$~dex in difference,
with the exception of the highest redshift bin, which is almost compatible with the luminosity of main-sequence galaxies. The resulting relation with stellar mass is flatter than the relation observed for star-forming galaxies. The relation is limited to just a few (high-) mass bins because in the mass regime below 10$^{10}$ M$_{\odot}$ , the X-ray flux was not high enough to constrain the X-ray luminosity of the subsample: we obtained a 95th percentile luminosity value $<0$. 
According to the hardness ratio analysis performed by \citet{2016ApJ...823..112P} on a subsample from C-COSMOS of early-type stacked galaxies, the X-ray luminosities of our quiescent galaxies are expected to be compatible with a combination of thermal and AGN emission in the lower two redshift bins and with highly obscured AGN at higher redshifts. This confirms the origin of the emission.

In Figure \ref{fig:L_X} we compare our results for the sample of star-forming galaxies with literature results by \citet{2012ApJ...753L..30M}, \citet{2015ApJ...800L..10R}, and \citet{2018MNRAS.474.1225A} after scaling the luminosity to the same $k$-correction as we used in our analysis where necessary (i.e., rescaled to the same photon index $\Gamma$ in the $k$-correction). At $z\sim1,$ our data points agree well with those of \citet{2012ApJ...753L..30M} and \citet{2018MNRAS.474.1225A}.
At $z\sim2$ our results are in great agreement, within the errors, with the results of \citet{2018MNRAS.474.1225A}, while our X-ray luminosities are systematically lower than those by \citet{2012ApJ...753L..30M} and \citet{2015ApJ...800L..10R}.
\begin{table*}
\caption{Fit parameters of the relations in Fig.~\ref{fig:SF_BH_all}. The equations we used are for $\log_{10}{\rm SFR}=m\,\log_{10}{\rm M}_*+q$; for $\log_{10}{\rm BHAR}=m\,\log_{10}{\rm M}_*+q$; and for the ratio $\log\frac{\text{BHAR}}{\text{SFR}}=\alpha\,\log\text{M}_*+\beta$.}             
\label{table:all_fitpar}      
\centering                          
\begin{tabular}{l | c c | c c | c c }        
\hline\hline                 
              & SFR                 &                  & BHAR              &                   & BHAR/SFR          &                    \\
Redshift bin  & $m $                & $q$              & $m$               & $q$               & $\alpha$          & $\beta$            \\    
\hline                                                                                                           
\textbf{Star-forming} &              &                  &                   &                   &                   &       \\
$0.10<z<0.65$ & $ 0.71  \pm 0.02  $ & $ -6.4 \pm 0.2 $ & $ 1.02 \pm 0.03 $ & $-13.7 \pm 0.3$ & $ 0.72 \pm 0.09 $ & $ -11.6  \pm 0.9  $ \\    
$0.65<z<1.30$ & $ 0.649 \pm 0.013 $ & $ -5.49\pm 0.13$ & $ 0.87 \pm 0.02 $ & $-11.5 \pm 0.2$ & $ 0.58 \pm 0.03 $ & $  -9.9  \pm 0.4  $ \\
$1.30<z<2.25$ & $ 0.639 \pm 0.018 $ & $ -5.06\pm 0.19$ & $ 1.18 \pm 0.03 $ & $-14.4 \pm 0.3$ & $ 0.65 \pm 0.04 $ & $ -10.6  \pm 0.4  $ \\
$2.25<z<3.50$ & $ 0.60  \pm 0.04  $ & $ -4.3 \pm 0.4 $ & $ 1.00 \pm 0.05 $ & $-12.2 \pm 0.5$ & $ 0.53 \pm 0.07 $ & $  -9.3  \pm 0.7  $ \\
\hline
\textbf{Quiescent}  &               &                &                   &                   &                   &                     \\
$0.10<z<0.65$     & $ 0.40\pm 0.14$ & $-4.5 \pm 1.5$ & $ 0.58\pm 0.17$ & $ -9.9 \pm 1.9 $  &                   &                     \\    
$0.65<z<1.30$     & $ 0.37\pm 0.15$ & $-3.6 \pm 1.6$ & $ 0.41\pm 0.16$ & $ -7.4 \pm 1.7 $  &   -               & $ -3.38 \pm 0.07 $  \\
$1.30<z<2.25$     & $-0.2 \pm 0.5 $ & $ 3.1 \pm 4.9$ & $ 0.5 \pm 0.2 $ & $ -7.6 \pm 2.4 $  &                   &                     \\
$2.25<z<3.50$     & -               & -              & $ 0.59\pm 0.18$ & $ -7.9 \pm 2.0 $  &                   &                     \\
\hline
\textbf{Starburst}  &              &                  &                   &                   &                   &                    \\
$0.10<z<0.65$ & $ 0.73 \pm 0.05 $  & $ -5.8 \pm 0.5 $ & $ 0.86 \pm 0.09 $ & $ -11.0 \pm 1.0 $ & $ 0.36 \pm 0.12 $ & $ -7.9 \pm 1.3 $  \\  
$0.65<z<1.30$ & $ 0.61 \pm 0.03 $  & $ -4.3 \pm 0.3 $ & $ 0.86 \pm 0.10 $ & $ -11.1 \pm 1.1 $ & $ 0.22 \pm 0.11 $ & $ -6.7 \pm 1.2 $  \\
$1.30<z<2.25$ & $ 0.54 \pm 0.07 $  & $ -3.1 \pm 0.8 $ & $ 1.08 \pm 0.11 $ & $ -12.8 \pm 1.2 $ & $ 0.54 \pm 0.15 $ & $ -9.6 \pm 1.6 $  \\
$2.25<z<3.50$ & $ 0.59 \pm 0.14 $  & $ -3.4 \pm 1.5 $ & $ 1.0  \pm 0.7  $ & $ -12.5 \pm 7.3 $ & $ 0.44 \pm 0.72 $ & $ -9.2 \pm 7.8 $  \\
\hline                                   
\end{tabular}
\end{table*}

\section{Constraining the coevolution of galaxy and black hole accretion} \label{sec:BH_SF}
In Figure~\ref{fig:SF_BH_all} we compare the SFR and the BHAR as a function of M$_*$ for the three categories of galaxies.
In the left column we show star-forming galaxies, in the middle column we list quiescent galaxies, and the right column contains starburst galaxies. 
The top panel shows the total SFR$_{\rm UV+IR}$ (SFR$_{\rm IR}$ only for starburst galaxies, see Sec.~\ref{ssec:SFR}) for all the redshift bins, the middle panel shows the BHAR of the sample obtained from the X-ray luminosity (Figure~\ref{fig:L_X}) using Equation \ref{eq:BHAR}, and the bottom panel presents the ratio between BHAR and SFR. 
We show linear fits in all panels except for the BHAR/SFR of quiescent galaxies where we report the average value of all redshifts and mass bins because no evolutionary trends are apparently visible.
The best-fit parameters are shown in Table \ref{table:all_fitpar}.

\subsection{Star-forming galaxies}
        For star-forming galaxies, the top panel shows the well-known evolution of the main-sequence of star-forming galaxies with time: the SFR follows a sublinear relation (in logaritmic units) with stellar mass that evolves in normalization with redshift, maintaining an almost constant slope and a bend at high M$_*$ in the two lower redshift bins. This bend was introduced by including the quiescent $24\mu m$ MIPS/\textit{Spitzer} detected sources, which as described above, are concentrated at high masses and lower redshifts. The median SFR$_{\rm UV}$ of the sample is  lower than the median SFR$_{\rm IR}$ by a factor that varies from 3 (low mass) to 100 (high mass, see Table~\ref{tab:SF_prop}), which means that the SFR$_{\rm UV}$ is negligible in most cases.

The observed BHAR distribution has a higher slope in the log-log plane; it is slightly superlinear 
(a direct consequence of the L$_\text{X}$-M$_*$ relation). 
    It is present at all considered redshifts and evolves with it.
    We note that the decrease in BHAR normalization evolves faster in the lowest redshift bin, but maintains a more constant increase at earlier cosmic epochs. The total decrease in BHAR at given M$_*$ is about 1.5~dex, and in SFR, the decrease is about 1-1.2~dex.

The ratio of BHAR to SFR provides interesting insights into how these two phenomena relate to each other. It increases with mass with a slope that varies in the $\sim0.5-0.7$ range, decreases at higher redshifts, and does not evolve in normalization. 
The positive slope of the BHAR/SFR can be interpreted as 
an indication of
an increased accretion efficiency in galaxies with high stellar mass by driving the infalling gas directly toward their inner core, which fuels a faster accretion onto the black hole. This can also be considered to mean that the main driver of the black hole accretion is not time, but the initial stellar mass. 
We note that the analogy between the increase in the BHAR/SFR as a function of stellar mass and the increase in the density of the central kiloparsec of the galaxy $\Sigma_1$ with mass \citep{2013ApJ...776...63F,2017ApJ...840...47B}. It has been proposed that more massive galaxies might have denser cores that might allow for a faster growth of the black hole.

\subsection{Quiescent galaxies}
In the central column of Fig.~\ref{fig:SF_BH_all} we show the data points of the quiescent galaxies. As expected, the SFR of these galaxies is lower than that of star-forming galaxies.
The two lower redshifts show a weak dependence of stellar mass, with a slope of $\sim0.4$, 
and the evolution with redshift in the bins in which we were able to constrain the SFR is also clear.
The reason might be that high-redshift elliptical galaxies are of a different nature than local galaxies and have higher percentages of gas \citep{2018NatAs...2..239G}, but part of the emission might also originate in dust cirrus illuminated by older stellar populations \citep{2007ApJ...657..810D, 2015A&A...573A.113B}. 
Another factor that may be contributing to the evolution we see is a small fraction of misclassified star-forming galaxies at high redshift. This may be caused by galaxies that emit in the MIPS band but lie below the detection threshold or by cross-contamination in the NUV$ - r / r - J$ because the uncertainty in the estimation of the rest-frame magnitudes that are required to classify a galaxy in the color-color diagram is higher. Overall, it is not possible to determine whether this evolution in time and mass dependence of quiescent galaxies and their mass dependence is significant, especially because we are able to constrain only these high-mass bins.

 For comparison with star-forming galaxies, we note that the median SFR$_{\rm UV}$ of the quiescent sample is lower by a factor of about 15 than the median SFR$_{\rm IR}$ $ \text{}$ (see Table~\ref{tab:Q_prop}). This means that the SFR$_{\rm UV}$ is negligible in most cases.

We are better able to constrain the BHAR at all redshifts and see a redshift evolution and weak dependence on stellar mass,with a slope $\sim0.5,$ even though it is difficult to analyze the slope evolution because of the small number of data points.
The normalization in BHAR is higher than in SFR, that is, the normalization of the BHAR is similar to the normalization of BHAR in star-forming galaxies, while the SFR of quiescents is clearly lower than that of star-forming galaxies. This indicates that the efficiency in accreting material onto the black hole is higher than in forming stars in galaxies at this late stage in the life of galaxies: the gas present in the galaxy ``prefers'' to fall into the black hole rather than form stars. This is consistent with the results of \citet{2018NatAs...2..239G}, who reported that a substantial amount of gas is available at high redshift in quiescent galaxies. This is then probably consumed less efficiently than in star-forming galaxies.
    
    The BHAR/SFR of quiescent galaxies, for the few mass and redshift bins where it was possible to constrain this, shows a flat trend in mass that is compatible with a constant value of $4.2\times10^{-4}$ at all redshifts and M$_*$. This value, obtained as a weighted mean of the data points, is a confirmation of the trends we have seen for BHAR and SFR: it is compatible with the values obtained for the highest mass bin of star-forming galaxies, 
    indicating that even though the ability of a galaxy to form stars has decreased in these galaxies, the efficiency of attracting the gas to the black hole did not.

\subsection{Starburst galaxies}
In the right column of Figure~\ref{fig:SF_BH_all} we show the starburst sample. By selection, FIR-selected starbursts have higher SFR than normal star-forming galaxies; the evolution in time is clear, as is the dependence on M$_*$.  

In starburst galaxies, the BHAR shows a close to linear dependence on mass and a weak evolution in redshift: values range in about $\text{}1$~dex, but the highest redshift bin has a lower normalization than the previous one, differently from what is seen in the other phases of galaxy life. 
This decrease in BHAR, which is not as pronounced as the decrease in star-forming galaxies (about $\sim1.5$~dex) causes them to slowly become starburst galaxies (i.e., higher than main-sequence galaxies) in the black hole accretion as well at lower redshift. 
    
    There is no clear evolution in the normalization of the BHAR/SFR, but it is interesting to note that the highest redshift bin shows the lowest normalization, which indicates that in these extremely star-forming galaxies black hole accretion was disfavored at higher redshifts. The slope appears to be positive and of about $0.2-0.4$.
    
    We tested selecting a starburst galaxy sample using the main-sequence from \citet{2011ApJ...739L..40R} with a redshift evolution as in \citet{2012ApJ...747L..31S}, SFR$_{\rm MS}(z)\propto (1+z)^{2.8}$. This selection assumed no bend on the main-sequence at high stellar masses and was calibrated up to $z\sim 2$. When we extrapolated this evolution to higher redshifts, we found a higher normalization than was reported by \citet{2015A&A...575A..74S}, which instead was calibrated up to $z\sim5$. This starburst selection led to no significant differences in the SFR of the sample, but to slightly lower BHAR values. We were also unable to constrain the X-ray luminosities of starbursts in the highest redshift bin. This suggests that the most starbursty galaxies accrete their black hole even less. This finally resulted in a flatter trend of the BHAR/SFR against stellar mass.

\subsection{Comparisons with the literature} \label{sec:compare}
\begin{figure}
   \centering
   \includegraphics[trim={1.8cm 1.5cm 2.3cm 0.8cm}, clip, width=\columnwidth]{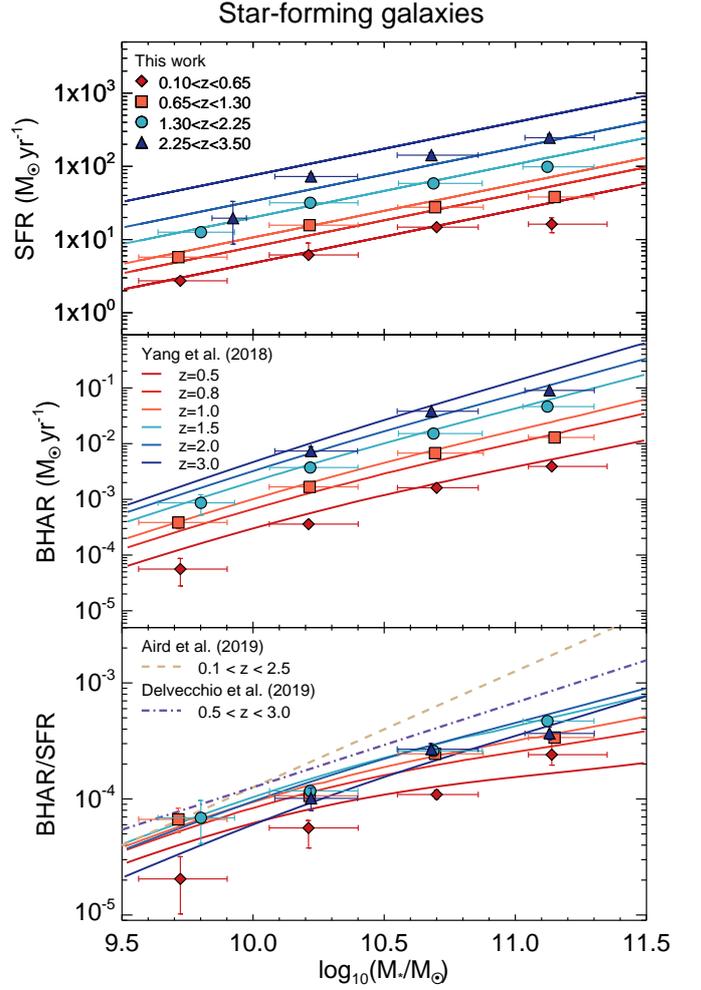}
      \caption{ Comparison between our results for star-forming galaxies and data from \citet{2018MNRAS.475.1887Y}, \citet{2019MNRAS.484.4360A}, and \citet{2019ApJ...885L..36D}.
            Data points are taken from the left column of Figure~\ref{fig:SF_BH_all}, and the curves are adapted from Figure~14 in \citet{2018MNRAS.475.1887Y} and were scaled to the same $k$-correction as we adopted here.
            Data from \citet{2019MNRAS.484.4360A} are taken from Fig.~13.
              }
         \label{fig:conf_yang}
   \end{figure}

In Figure~\ref{fig:conf_yang} we compare our results for the star-forming sample with results from the literature. We show our data points from 
Figure~\ref{fig:SF_BH_all} in the left column and compare them with results from \citet{2018MNRAS.475.1887Y} as solid lines, with results from \citet{2019MNRAS.484.4360A} as a dashed beige line, and with results from \citet{2019ApJ...885L..36D} as a dot-dashed violet line.

         Our data points qualitatively agree with those from \citet{2018MNRAS.475.1887Y} in the three panels
         . \citet{2018MNRAS.475.1887Y} selected their star-forming sample from GOODS-N, GOODS-S, and COSMOS UltraVISTA DR1 based on the SFR from SED fitting with a SFR threshold at $1.3$~dex below the main-sequence. SFRs are from \citet{2015ApJ...801...97S} and \citet{2019ApJS..243...22B}). Their average SFRs are slightly higher than ours. They also reported that the BHAR of the lowest redshift bin is separated more than the BHAR between the other bins.
         
The BHAR/SFR of \citet{2019MNRAS.484.4360A} has a linear trend with mass with a higher slope than ours. Their BHARs were obtained with the same method as in \citet{2018MNRAS.474.1225A}. Their SFRs are from SED fitting from the UV to MIR, which means that the FIR was not included. Together with the constant bolometric correction they used, this might lead to the higher slope value.
   
   We also compared our data points with the relation found by \citet{2019ApJ...885L..36D} through an empirically motivated model that successfully reproduces the observed X-ray luminosity function (XLF) since $z\sim3$. With this model, they found a growth in two steps: until the galaxy reaches a critical mass, the black hole growth lags behind it, and then, as the stellar mass increases, the BHAR is enhanced with respect to the SFR, following a superlinear relation very similar to ours, except for an offset of $\sim0.1-0.3$~dex.

\section{Comparison between the evolution of sBHAR and sSFR} \label{sec:specifics}
%
We computed the average sSFR (i.e., specific SFR, SFR/M$_*$) and the sBHAR (i.e., specific BHAR, BHAR/M$_\text{BH}$) for all the bins in M$_*$ and $z$. In this section we describe how the sBHAR was derived and we discuss the results.

   \begin{figure*}
   \centering
   \includegraphics[trim={0 4cm 0.5cm 5.1cm}, clip,width=0.8\paperwidth]{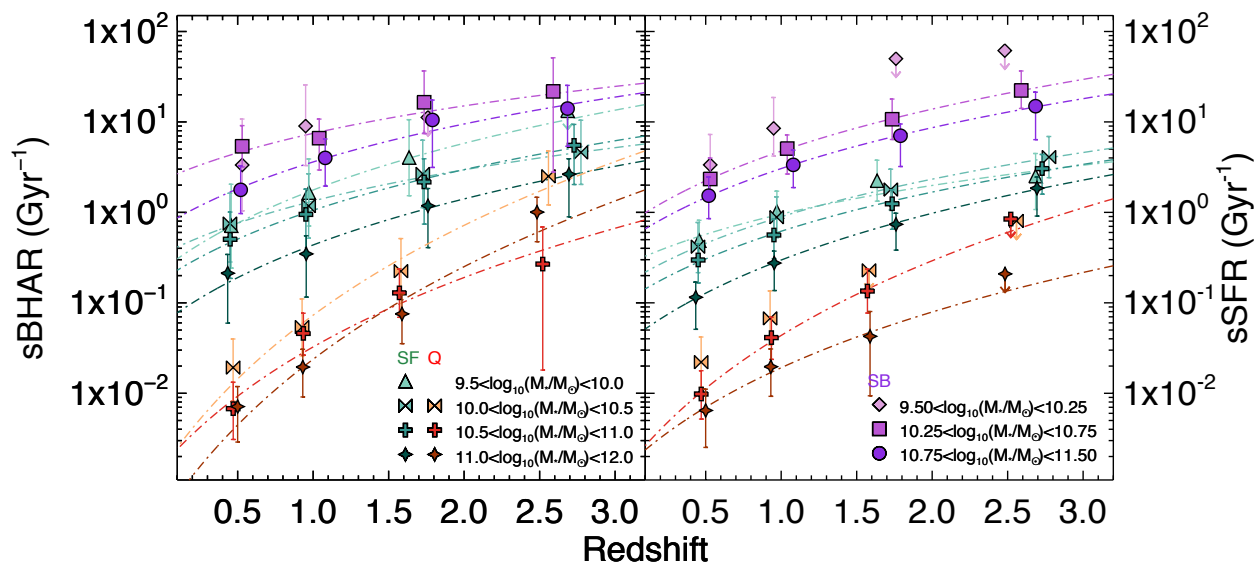}
      \caption{ Specific black hole accretion rate (left panel) and sSFR (right panel) as a function of redshift for star-forming (SF, in green), quiescent (Q, in red), and starburst (SB, in purple) galaxies. The data points are placed at the median redshift of each mass bin and are coded in shape and color according to their median mass and type. Error bars represent the 90\% confidence interval associated with each measure. We show an upper limit (the 95th percentile) for the data points whose sBHAR/sSFR 5th percentile is compatible with zero and for starburst galaxies at high redshift and low mass, where our sample is incomplete. The dot-dashed line is the best fit of the data to the curve $sBHAR(sSFR)=\delta\, (1+z)^{\gamma}$ and is color-coded according to the galaxy type and mass bin.
              }
         \label{fig:sBHAR_all}
   \end{figure*}
%
\subsection{Black hole mass estimate}
In order to estimate the black hole mass M$_{\text{BH}}$ , we considered the fits in the bottom panels of Fig. \ref{fig:SF_BH_all} and in the right column of Table~\ref{table:all_fitpar}, obtained from the equation
\begin{equation}  \label{eq:ratio_fit}
\centering
\log\frac{\text{BHAR}}{\text{SFR}}=\alpha\log\text{M}_*+\beta.
\end{equation}

The general equation we used for the computation is
\begin{equation}  \label{eq:ratio_comp}
\centering
\log\frac{\text{BHAR}}{\text{SFR}}=
\log\frac{\dot{\text{M}}_{\text{BH}}}{\dot{\text{M}}_*}=
\log\frac{\partial\text{M}_{\text{BH}}}{\partial\text{M}_*}= 
\alpha \log\text{M}_* + \beta,
\end{equation}
which is valid only under the assumption that the growth rates of stellar and black hole mass are constant in time. The ratios we measure show almost no evolution up to $z\sim3.5,$  which appears to justify this assumption.
By integrating Equation~\ref{eq:ratio_comp} with respect to stellar mass, we obtained the estimate of the M$_\text{BH}$ of each stellar mass and redshift bin:
\begin{equation}  \label{eq:M_bh}
\centering
\text{M}_\text{BH}=\frac{10^\beta}{1+\alpha}\,\text{M}_*^{1+\alpha},
\end{equation}
where $\alpha$ and $\beta$ are the same parameters in Table~\ref{table:all_fitpar} and are the parameters we used in the computation. It follows from Eq.~\ref{eq:M_bh}  that the black hole mass has a superlinear dependence on stellar mass in star-forming and starburst galaxies, but M$_{\rm BH}$ has a linear dependence on M$_*$ in quiescent galaxies.

If instead of performing an indefinite integral we integrate between the initial masses ($\text{M}_{*,i}$ and $\text{M}_{\text{BH},i}$) and the final (observed) masses, the integrated Equation~\ref{eq:ratio_comp} would read
\begin{equation}  \label{eq:M_bh_compl}
\centering
\text{M}_\text{BH}=\frac{10^\beta}{1+\alpha}\,(\text{M}_*^{1+\alpha}-\text{M}_{*,i}^{1+\alpha})+\text{M}_{\text{BH},i}
.\end{equation}
We assumed that the initial black hole mass is lower by at least an order of magnitude ($\lesssim10^5\text{M}_*$) than the final mass, however, therefore it is negligible. The initial stellar mass instead eludes us, and becuase we subtracted it, our $\text{M}_\text{BH}$ is an upper limit. We therefore decided not to use the complete expression of $\text{M}_\text{BH}$ in Eq.~\ref{eq:M_bh_compl}, but the expression in Eq.~\ref{eq:M_bh}.

\subsection{Results for the specific accretions}
Figure~\ref{fig:sBHAR_all} shows the redshift evolution of the sBHAR (estimated using $\text{M}_{\text{BH}}$ from Eq.~\ref{eq:M_bh}) and sSFR for star-forming, quiescent, and starburst galaxies.
sBHAR and sSFR have a decreasing trend toward lower redshift for the three galaxy types, and the stellar mass shows a split: low-mass galaxies have higher values than high-mass galaxies, although this split is often not significant within the error bars.

        These trends confirm what has been reported before about these specific accretions. They also  provide new interesting insight.
        The spread in stellar mass is consistent with downsizing: it implies that high-mass galaxies have accreted most of their M$_*$ and M$_\text{BH}$ at high redshift and their accretion decreased fast and steeply, whereas low-mass galaxies have accreted their mass more slowly, but   their accretion rate decreased more slowly with  time. Downsizing has previously been observed for stellar mass \citep{1996AJ....112..839C, 2006A&A...453L..29C} and black hole luminosity (e.g., \citealt{2004MNRAS.351..169M} but also \citealt{2004MNRAS.354.1020S, 2009ApJ...690...20S, 2015ApJ...810...74A}, who reached similar conclusions from continuity equation arguments). 
        In our plots, downsizing can be observed in the steepening of the slope $\gamma$ at higher masses in star-forming and starburst galaxies and in the two specific accretions, which indicate a faster decrease for higher masses. Downsizing can also be seen in
        high-mass galaxies (darker data points), which show lower sSFR(sBHAR) on average than lower mass galaxies. This means that at even at higher redshifts, the most massive galaxies have already accreted most of their stellar and black hole mass, even though within the error bars, sSFR (sBHAR) data points are often compatible with a unique value for all stellar masses at a given redshift. In particular, this trend would not be present in star-forming galaxies without introducing the dependence of the black hole mass on the slope $\alpha$ from Equation~\ref{eq:ratio_fit}.

We show in Figure~\ref{fig:sBHAR_all} the best fits to the data when a redshift evolution of the form
sBHAR~(sSFR)$=\delta\, (1-z)^{\gamma}$ is adopted, which we applied to each mass bin for all galaxy types when at least three data points were available\footnote{Fits performed with IDL/MPFIT \citep{2009ASPC..411..251M}}. 
The sSFR and sBHAR for star-forming and starburst galaxies are compatible with $\gamma=2.8,$ as in \citet{2012ApJ...747L..31S}, within $1\sigma$, but this is not the case for quiescent galaxies, which have higher slopes with a steeper decreases in specific accretions in time. Interestingly, we do not note a significant exponent difference between sBHAR and sSFR at given galaxy type.
When we consider that there may be a small contribution from misclassified star-forming galaxies in the quiescent sample, especially at high redshift, the real trend may be even flatter, and approach a slope $\gamma=2.8$.
We do see a difference in the normalizations of the relations, which show lower specific accretions for quiescent galaxies, followed by star-forming galaxies, and finally, starburst galaxies, which tend to have higher specific accretions.
When  the normalizations for the two specific accretions are compared, they are very similar in the case of starburst galaxies, while star-forming galaxies have a higher normalization for the sBHAR and quiescent galaxies show higher sBHAR than sSFR at high $z$.

    Because the inverse of sSFR and sBHAR can be considered as the mass-doubling timescale of the M$_*$ and of the SMBH, this means that the stellar mass of the galaxy and the SMBH are accreted faster in starburst and star-forming galaxies, which especially at high redshift are still efficient and can quickly double their mass. Quiescent galaxies are instead slower and have mass-doubling timescales of about a Hubble time or more. The black hole mass-doubling time of star-forming galaxies instead seems to be shorter than the stellar mass black hole mass-doubling time at every redshift and mass bin.
    
These similar evolution trends between sBHAR and sSFR strongly suggest a connection between the two accretions that appears to be present in all galaxy life phases. Star-forming galaxies appear to dominate the accretion histories; they are the most numerous galaxies and are able to substantially accrete their stellar and black hole masses. Starburst galaxies, with a higher capability of accretion but short-lived episodes, and quiescent galaxies, although their accretion capabilities are lower, appear to be able to accrete their black hole more efficiently than their stellar mass.
   
\begin{table*}
\caption{Fit parameters of the relations in Fig.~\ref{fig:sBHAR_all}. The equation we used is sBHAR~(sSFR)$=\delta\, (1 +z)^{\gamma}$.}             
\label{table:specific_fitpar}      
\centering                          
\begin{tabular}{l | c c | c c }        
\hline\hline                 
Mass bins                   & sBHAR parameters      &                 & sSFR parameters     &                 \\
$\log_{10}($M$_*$/M$_\sun)$ & $\delta$              &  $\gamma$       & $\delta$            &  $\gamma$       \\
\hline                                                                                                  
\textbf{Star-forming}       &                       &                 &                     &                 \\
 9.5--10.0                  & $ 0.2  \pm 0.3  $     & $ 2.9 \pm 1.8 $ & $ 0.28 \pm 0.13 $   & $ 1.8 \pm 0.5 $ \\
10.0--10.5                  & $ 0.3  \pm 0.3  $     & $ 2.0 \pm 0.9 $ & $ 0.17 \pm 0.09 $   & $ 2.4 \pm 0.6 $ \\
10.5--11.0                  & $ 0.18 \pm 0.14 $     & $ 2.6 \pm 0.8 $ & $ 0.11 \pm 0.05 $   & $ 2.5 \pm 0.5 $ \\
11.0--12.0                  & $ 0.06 \pm 0.04 $     & $ 2.9 \pm 0.7 $ & $ 0.039\pm 0.016$   & $ 2.9 \pm 0.4 $ \\
\hline
\textbf{Quiescent}          &                       &                 &                     &                 \\
10.0--10.5                  & $ 0.0015 \pm 0.0012 $ & $ 5.6 \pm 0.9 $ & -                   & -               \\
10.5--11.0                  & $ 0.0016 \pm 0.0009 $ & $ 4.4 \pm 0.7 $ & $0.0017 \pm 0.0012$ & $ 4.7 \pm 1.0 $ \\
11.0--12.0                  & $ 0.0004 \pm 0.0003 $ & $ 5.8 \pm 0.8 $ & $0.0017 \pm 0.0013$ & $ 3.5 \pm 1.1 $ \\
\hline                      
\textbf{Starburst}          &                       &                 &                     &                 \\
 9.50--10.25                & -                     & -               & -                   & -               \\
10.25--10.75                & $ 2.3   \pm 1.6 $     & $ 1.7 \pm 0.9 $ & $ 0.8  \pm 0.4 $    & $ 2.6 \pm 0.5 $ \\
10.75--11.50                & $ 0.7   \pm 0.4 $     & $ 2.4 \pm 0.6 $ & $ 0.5  \pm 0.2 $    & $ 2.6 \pm 0.5 $ \\
\hline
\end{tabular}
\end{table*}
   
   \begin{figure}
   \centering
   \includegraphics[trim={2cm 1.5cm 3cm 2cm}, clip, width=\columnwidth]{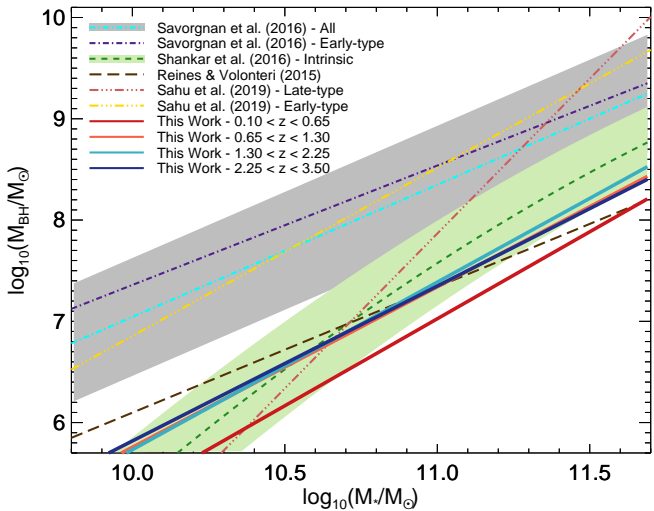}
      \caption{ Comparison between our M$_*$ and M$_\text{BH}$ from Eq.~\ref{eq:M_bh} and parameters $\alpha$ and $\beta$ from Table~\ref{table:all_fitpar} for star-forming galaxies and results from models in the literature. Data from this work are shown as solid lines and are color-coded according to redshift as in Fig.~\ref{fig:SF_BH_all} and \ref{fig:conf_yang}.
      The relations from \citet{2016ApJS..222...10S} use the M$_*$ of the bulge.
      The gray shaded area is the confidence range for \citet{2016ApJS..222...10S}, and the light green are is the confidence range for \citet{2016MNRAS.460.3119S}.
              }
         \label{fig:comp_models}
   \end{figure}


\section{Relation between stellar mass and black hole mass} \label{sec:M_vs_M}
In Figure~\ref{fig:comp_models} we show our best-fit resulting M$_\text{BH}$-M$_*$ relations for star-forming galaxies derived from Eq.~\ref{eq:M_bh} with parameters from Table~\ref{table:all_fitpar}. These are superlinear relations with approximately M$_{\rm BH}\propto\text{M}_*^{1.6}$.
Our stellar masses are to be considered lower limits because as pointed out in \citet{2003ApJ...585L.117B}, stellar masses obtained using \citet{2003MNRAS.344.1000B} models lead to lower mass-to-light ratios and therefore to systematically lower stellar mass estimates at a given luminosity. A possible correction in our stellar masses that would cause them to agree better with those of \citet{2003ApJ...585L.117B} (increase of $\sim0.25$~dex) would further support our results.

We note that recent scaling relations proposed by other groups for both early-type galaxies and late-type galaxies show different scaling slopes and normalizations. \citet{2019ApJ...876..155S} find a relation for early-type galaxies that is similar in both slope and normalization to the lines of \citeauthor{2016ApJS..222...10S} (who used bulge stellar masses. See \citet{2016MNRAS.460.3119S, 2019MNRAS.485.1278S} for details) reported in Fig. \ref{fig:comp_models}. For late-type galaxies, \citet{2019ApJ...876..155S} find a lower normalization, more consistent with our relation around M$_*\sim 10^{10.5}{\rm M}_\odot$, but a much steeper slope of $\sim3.0\pm 0.5$, about $3\sigma$ steeper than ours. All in all our results point to either lower normalizations or flatter slopes than those identified from local dynamically measured SMBHs.
On the other hand, our resulting scaling relations have a similar normalization with those proposed by \citet{2015ApJ...813...82R} from local broad line AGN, despite them finding a close to linear relation, and agree well with 
the intrinsic scaling relation by \citet{2016MNRAS.460.3119S}. The latter have suggested that the local sample of (mainly) early-type galaxies with dynamically measured SMBH may be biased because a preselection might favor galaxies with the highest black hole masses and related gravitational radii. As recently suggested by \citet{2019MNRAS.485.1278S}, AGN samples that are clearly not affected by resolution-related selections effects should be closer to the intrinsic scaling relations. Our study further supports this view. 

We note that our normalization is inversely proportional to the radiative efficiency assumed in Eq.~\ref{eq:BHAR}. A match to the normalization of the local  M$_\text{BH}$-M$_*$ relation by \citeauthor{2016ApJS..222...10S} would require mean radiative efficiencies that are an order of magnitude lower. This is disfavored on other grounds \citep{2020MNRAS.493.1500S}. 

Our analysis adds key evidence, using IR data, to an increasing body of work \citep{2019ApJ...885L..36D,  2020ApJ...888...37D, 2020MNRAS.493.1500S, 2020ApJ...889...32S} on the weak evolution in the BHAR/SFR and its relatively low normalization relative to local raw dynamical M$_\text{BH}$-M$_*$ relations.

\section{Conclusions} \label{sec:conclusions}
We performed a statistical study on the COSMOS field in order to constrain the history and coevolution of star formation and black hole accretion in star-forming, quiescent, and starburst galaxies. We selected a mass-complete sample from the COSMOS 2015 catalog \citep{2016ApJS..224...24L} by classifying normal star-forming and quiescent galaxies through the NUV$ - r / r - J$ color-color diagram, and we selected starburst galaxies from the \textit{Herschel}-selected sample in \citet{2013MNRAS.432...23G}. We performed an X-ray stacking analysis and combined it with detected sources from \citet{2016ApJ...819...62C} in order to estimate the average X-ray luminosity and therefore average BHAR. We estimated the SFR from FIR stacking and UV SED fitting. Our main results are listed below.

   \begin{enumerate}
      \item We find a robust L$_{\rm X} - {\rm M}_*$ relation for star-forming galaxies at all considered redshifts. This relation evolves with an increasing normalization at higher redshifts. The X-ray luminosity of quiescent galaxies is close to that of star-forming galaxies, especially at low masses. At high masses, L$_\text{X}$ in quiescent galaxies is lower than in star-forming galaxies. The X-ray luminosities of starburst galaxies are compatible with star-forming at high redshifts and evolve mildly down to low redshift, where they are clearly higher than those of star-forming galaxies.
      \item The L$_{\rm X} - {\rm M}_*$ relation translates into a BHAR-M$_*$ relation in Fig.~\ref{fig:SF_BH_all} (middle row) that shows that the evolution of the BHAR in star-forming galaxies is faster at lower redshifts, it has a pronounced redshift evolution, and a weak mass dependence in quiescent galaxies. In turn, starburst galaxies have a marked mass dependence and a distinctive redshift evolution: going back in time, it reaches a maximum at $z\sim 1.7$ to then decrease again.
      \item BHAR in star-forming galaxies increases more with stellar mass than the SFR. In quiescent galaxies, the BHAR values lie close to the BHAR of star-forming galaxies, while the SFR of quiescent galaxies is clearly below the main-sequence. It is interesting that while the SFR of starburst galaxies continues to increase at higher redshifts, at the highest redshift in our study, the BHAR of these galaxies decreases.
      \item The ratio between BHAR and SFR in star-forming and starburst galaxies has a positive relation M$_*$ that is almost time-independent. The ratio is higher for quiescent galaxies, compatible with a flat trend in M$_*$ , indicating indicating a stronger tendency for this type of galaxy to accrete onto the black hole than to form stars, regardless of stellar mass. From this it follows that M$_{\rm BH}$ has a superlinear dependence on M$_*$ in star-forming and starburst galaxies, and the dependence is linear in quiescent galaxies.
      \item sBHAR and sSFR follow very similar decreasing trends in time
      . We see signs of downsizing in all types of galaxies, a faster accretion (of M$_{\rm BH}$ and M$_*$) in starbursts followed by star-forming galaxies, and finally, by quiescent galaxies with mass-doubling timescales of about a Hubble time. 
      \item The resulting M$_\text{BH}$-M$_*$ relation from our data agrees well with independent determinations of the relation that were retrieved from AGN samples and Monte Carlo simulations.
   \end{enumerate}
   All of these results confirm the coevolution of host galaxy and black hole follows the pattern of downsizing at all redshifts and in different galaxy evolutionary phases. In this picture, the bulk of the black hole and stellar masses is accreted in galaxies during the main-sequence phase through secular processes, where more massive galaxies are more efficient at accreting the black hole. Starburst episodes play a lesser role for both accretions because only a few galaxies are in this phase and these episodes are only weakly able to enhance black hole accretion at high redshifts. 
   The deeper potential well of more massive and possibly more compact galaxies seems to be playing a role in feeding the black hole more efficiently in star-forming and starburst galaxies.
   In the quiescent life phase of galaxies, the black hole accretion is not as penalized as the star formation. The gas availability reported by \citet{2018NatAs...2..239G} means that this gas may not go to star formation because of different galactic properties in the different life phases (e.g., disk and bulge dynamics), but to accrete the black hole.
   Finally, we find additional evidence that suggests that the M$_\text{BH}$-M$_*$ may have a lower normalization than the local dynamical relation.

\begin{acknowledgements}
The authors want to thank the anonymous referee for his/her comments that clearly improved the consistency and overall quality of this paper.
    We acknowledge helpful conversations with Patricia Ar\'evalo, Francesca Civano, Francesca M. Fornasini, Mauro Giavalisco, Alister Graham and Claudio Ricci.
      RC acknowledges financial support from CONICYT Doctorado Nacional N$^\circ$\,21161487 and CONICYT PIA ACT172033; 
      RC and PC acknowledge support from the CONICYT/FONDECYT program N$^\circ$\,1150216;
      GR and CM acknowledge support from an INAF PRIN-SKA 2017 grant 1.05.01.88.04 and from the STARS@UniPD grant;
      PC acknowledges support from the BIRD 2018 research grant from the Universit\`a degli Studi di Padova;
      ID is supported by the European Union's Horizon 2020 research and innovation program under the Marie Sk\l{}odowska-Curie grant agreement N$^\circ$\,78867;
      FS acknowledges partial support from a Leverhulme Trust Research Fellowship.
\end{acknowledgements}

%
%

\bibliographystyle{aa} 
\bibliography{bibliografia}

\begin{appendix}
\section{Sample properties} \label{appendix}
We report the tables with additional properties of our sample. Table~\ref{tab:SF_prop} lists star-forming galaxies, Table~\ref{tab:Q_prop}  quiescent galaxies, and Table~\ref{tab:SB_prop} starbursts.

\begin{table*}\centering
\caption{Properties of the star-forming galaxy sample. For each mass and redshift bin we show the median mass and redshift, and the number of stacked and detected galaxies in the 2-7~keV band.
We also show the X-ray luminosity, the SFR estimated from the FIR, and the obscured SFR from the UV. Quantities are medians, and confidence ranges are 1$\sigma$.}
\vspace{5pt}
\label{tab:SF_prop}
\begin{tabular}{lc|c|c c|c c c}
\hline
\hline
$\text{Mass range}$                             & $M_*$ &  $z$  & N$_{\text{stacked}}$  & N$_{\text{detected}}$ & $L_{\text{2-10~keV}}$   & $\text{SFR}_{IR}$ & $\text{SFR}_{UV}$  \\     
$\log_{10}($M$_*$/M$_\sun)$     & $\log_{10}($M$_*$/M$_\sun)$   & &     &                       & $10^{42}\um{erg}\ump{s}{-1}$    & M$_\odot \ump{yr}{-1}$        & M$_\odot \ump{yr}{-1}$\\ 
\hline
\hline
\multicolumn{8}{l}{$0.10<z<0.65$}       \\
\hline
9.5--10.0               &  9.72    & 0.46       & 4244  & 19    & $0.023^{+0.007}_{-0.007}$     & $ 2.22^{+0.08}_{-0.07}$ & $0.500^{+0.012}_{-0.011}$             \T \B \\    
10.0--10.5          & 10.21    & 0.45   & 2514  & 75    & $0.142^{+0.009}_{-0.009}$     & $ 5.68^{+2.67}_{-0.17}$ & $0.479^{+0.013}_{-0.018}$             \T \B \\     
10.5--11.0              & 10.70    & 0.45       & 1214  & 117   & $0.639^{+0.015}_{-0.015}$     & $14.2 ^{+0.7 }_{-0.7 }$ & $0.56 ^{+0.03 }_{-0.02 }$             \T \B \\      
11.0--12.0          & 11.14    & 0.43   & 239   & 38    & $1.550^{+0.038}_{-0.036}$     & $15.7 ^{+2.2 }_{-2.4 }$ & $0.76 ^{+0.07 }_{-0.11 }$             \T \B \\       
\hline
\hline
\multicolumn{8}{l}{$0.65<z<1.30$}        \\
\hline
9.5--10.0               &  9.72   & 0.97        & 15431 & 56    & $ 0.15^{+0.02}_{-0.02} $       & $ 4.70^{+0.14}_{-0.13}$       & $1.051^{+0.009}_{-0.011} $    \T \B \\   
10.0--10.5          & 10.22   & 0.97    & 7957  & 175   & $ 0.66^{+0.03}_{-0.03} $       & $14.9 ^{+0.3 }_{-0.3 }$       & $0.854^{+0.018}_{-0.015} $    \T \B \\   
10.5--11.0              & 10.69   & 0.95        & 3888  & 370   & $ 2.67^{+0.05}_{-0.05} $       & $26.8 ^{+0.5 }_{-0.5 }$       & $0.856^{+0.023}_{-0.016} $    \T \B \\   
11.0--12.0          & 11.15   & 0.95    & 761   & 142   & $ 5.13^{+0.12}_{-0.12} $       & $37.1 ^{+2.7 }_{-2.1 }$       & $1.29 ^{+0.08 }_{-0.13 } $    \T \B \\   
\hline
\hline
\multicolumn{8}{l}{$1.30<z<2.25$}        \\
\hline                  
9.5--10.0               &  9.80    & 1.64       & 13451 & 47    & $ 0.35^{+0.09}_{-0.08} $       & $10.4^{+0.6}_{-0.6}$          & $2.17^{+0.02 }_{-0.02} $      \T \B \\   
10.0--10.5          & 10.22    & 1.73   & 11334 & 177   & $ 1.47^{+0.09}_{-0.10} $       & $30.2^{+0.7}_{-0.7}$          & $1.62^{+0.03 }_{-0.03} $      \T \B \\   
10.5--11.0              & 10.69    & 1.74       & 5273  & 368   & $ 5.97^{+0.17}_{-0.16} $       & $57.0^{+1.5}_{-1.5}$          & $1.18^{+0.02 }_{-0.02} $      \T \B \\   
11.0--12.0          & 11.12    & 1.76   & 954   & 172   & $18.21^{+0.43}_{-0.45} $       & $97.0^{+4.6}_{-4.7}$          & $1.46^{+0.04 }_{-0.06} $      \T \B \\   
\hline
\hline
\multicolumn{8}{l}{$2.25<z<3.50$}        \\
\hline
9.5--10.0               &  9.92    & 2.69       & 1264  & 3         & $  0.8^{+0.8 }_{-0.9} $    & $ 14.9^{+ 8.6}_{- 6.7}$       & $4.72^{+0.15}_{-0.17} $        \T \B \\  
10.0--10.5          & 10.22        & 2.77       & 7159  & 98    & $  2.9^{+0.4 }_{-0.4} $      & $ 68.2^{+ 4.4}_{- 4.1}$       & $4.49^{+0.07}_{-0.08} $        \T \B \\  
10.5--11.0              & 10.68    & 2.73       & 2726  & 166   & $ 14.9^{+0.6 }_{-0.6} $   & $139.4^{+ 9.1}_{- 9.1}$  & $2.78^{+0.06}_{-0.05} $        \T \B \\  
11.0--12.0          & 11.13        & 2.69       & 466   & 70    & $ 35.5^{+1.6 }_{-1.5} $   & $243.4^{+16.3}_{-14.8}$  & $2.44^{+0.12}_{-0.14} $        \T \B \\  
\hline

\end{tabular}
\end{table*}
\begin{table*}\centering
\caption{Same as Table~\ref{tab:SF_prop} for the quiescent galaxy sample.}
\vspace{5pt}
\label{tab:Q_prop}
\begin{tabular}{lc|c|c c|c c c}
\hline
\hline
$\text{Mass range}$                             & $M_*$ &  $z$  & N$_{\text{stacked}}$  & N$_{\text{detected}}$ & $L_{\text{2-10~keV}}$   & $\text{SFR}_{IR}$ & $\text{SFR}_{UV}$  \\     
$\log_{10}($M$_*$/M$_\sun)$     & $\log_{10}($M$_*$/M$_\sun)$   & &     &                       & $10^{42}\um{erg}\ump{s}{-1}$    & M$_\odot \ump{yr}{-1}$        & M$_\odot \ump{yr}{-1}$\\ 
\hline
\hline
\multicolumn{8}{l}{$0.10<z<0.65$}       \\
\hline
10.0--10.5          & 10.27    & 0.47   & 867   & 12    & $0.060^{+0.017}_{-0.018}$     & $0.40^{+0.07}_{-0.09}$  & $0.017^{+0.002}_{-0.002}$       \T \B \\ 
10.5--11.0              & 10.74    & 0.47       & 830   & 19    & $0.061^{+0.018}_{-0.015}$     & $0.51^{+0.12}_{-0.10}$  & $0.026^{+0.004}_{-0.004}$       \T \B \\ 
11.0--12.0          & 11.16    & 0.50   & 271   & 14    & $0.177^{+0.035}_{-0.035}$     & $0.96^{+0.21}_{-0.22}$  & $0.074^{+0.010}_{-0.017}$       \T \B \\ 
\hline
\hline
\multicolumn{8}{l}{$0.65<z<1.30$}        \\
\hline
10.0--10.5          & 10.30   & 0.93    & 2594  & 19    & $0.18^{+0.04}_{-0.05}$        & $1.3 ^{+0.3 }_{-0.4 }$  & $0.083^{+0.002}_{-0.002}$     \T \B \\   
10.5--11.0              & 10.75   & 0.94        & 3354  & 72    & $0.41^{+0.04}_{-0.04}$        & $2.2 ^{+0.3 }_{-0.2 }$  & $0.130^{+0.003}_{-0.004}$     \T \B \\   
11.0--12.0          & 11.15   & 0.93    & 1107  & 25    & $0.47^{+0.08}_{-0.08}$        & $2.8 ^{+0.5 }_{-0.4 }$  & $0.172^{+0.009}_{-0.010}$     \T \B \\   
\hline
\hline
\multicolumn{8}{l}{$1.30<z<2.25$}        \\
\hline                  
10.0--10.5          & 10.25    & 1.58   & 769   & 9     & $0.6^{+0.3}_{-0.3}$           & $- $                                            & $0.279^{+0.013}_{-0.011}$     \T \B \\   
10.5--11.0              & 10.75    & 1.57       & 1534  & 25    & $1.2^{+0.2}_{-0.2}$           & $7.4^{+0.8}_{-0.8}$             & $0.307^{+0.008}_{-0.008}$     \T \B \\   
11.0--12.0          & 11.14    & 1.59   & 532   & 8     & $1.8^{+0.4}_{-0.4}$           & $6.0^{+2.1}_{-3.0}$             & $0.400^{+0.021}_{-0.012}$     \T \B \\   
\hline
\hline
\multicolumn{8}{l}{$2.25<z<3.50$}        \\
\hline
10.0--10.5          & 10.32        & 2.56       & 105   & 3         & $ 8.4^{+2.5}_{-2.6}$          & $- $                                               & $1.25^{+0.22}_{-0.06}$        \T \B \\  
10.5--11.0              & 10.75    & 2.52       & 272   & 7     & $ 2.6^{+1.5}_{-1.5}$      & $- $                                             & $1.15^{+0.06}_{-0.07}$        \T \B \\  
11.0--12.0          & 11.12        & 2.48       & 66    & 6         & $22.2^{+3.6}_{-3.6}$      & $-$                                              & $1.71^{+0.18}_{-0.15}$        \T \B \\  
\hline
\end{tabular}
\end{table*}
\begin{table*}\centering
\caption{Same as Table~\ref{tab:SF_prop} for the starburst galaxy sample. We only use SFR$_{IR}$ for this sample.}
\vspace{5pt}
\label{tab:SB_prop}
\begin{tabular}{lc|c|c c|c c}
\hline
\hline
$\text{Mass range}$                             & $M_*$ &  $z$  & N$_{\text{stacked}}$  & N$_{\text{detected}}$ & $L_{\text{2-10~keV}}$   & $\text{SFR}_{IR}$  \\ 
$\log_{10}($M$_*$/M$_\sun)$     & $\log_{10}($M$_*$/M$_\sun)$   & &     &               & $10^{42}\um{erg}\ump{s}{-1}$    & M$_\odot \ump{yr}{-1}$        \\ 
\hline
\hline
\multicolumn{7}{l}{$0.10<z<0.65$}\\
\hline
9.50--10.25          &  9.87    & 0.53 & 58 & 4     & $0.36^{+0.09}_{-0.09}$  & $ 22^{+ 4}_{- 2}$    \T \B \\
10.25--10.75     & 10.52        & 0.53 & 6  & 5         & $4.68^{+0.40}_{-0.41}$  & $ 74^{+20}_{- 9}$    \T \B \\
10.75--11.50     & 10.96        & 0.52 & 2  & 3         & $6.64^{+0.67}_{-0.68}$  & $143^{+ 5}_{- 1}$    \T \B \\
\hline                                                                                     
\hline                                                                                     
\multicolumn{7}{l}{$0.65<z<1.30$} \\  
\hline                                 
9.50--10.25              &  9.90        & 0.95 & 162 & 7        & $0.8^{+0.2}_{-0.2}$    & $ 66^{+ 3}_{- 3}$     \T \B \\
10.25--10.75     & 10.42        & 1.04 & 161 & 10       & $2.5^{+0.3}_{-0.3}$          & $131^{+ 5}_{- 2}$    \T \B \\
10.75--11.50     & 10.91        & 1.08 & 43  & 5        & $6.2^{+0.6}_{-0.6}$    & $280^{+15}_{-16}$     \T \B \\
\hline
\hline
\multicolumn{7}{l}{$1.30<z<2.25$}\\
\hline                                                                                           
9.50--10.25              & 10.01        & 1.76 & 57  & 1        & $-$                    & $222^{+15}_{-18}$     \T \B \\
10.25--10.75     & 10.50        & 1.74 & 145 & 24       & $15.1^{+1.2}_{-1.2}$   & $331^{+16}_{- 9}$     \T \B \\
10.75--11.50     & 10.89        & 1.79 & 78  & 23       & $40.8^{+2.0}_{-2.0}$   & $536^{+34}_{-23}$     \T \B \\
\hline 
\hline
\multicolumn{7}{l}{$2.25<z<3.50$}\\  
\hline                                                                                   
9.50--10.25              & 10.07        & 2.48  & 5   & 0   & $-$                        & $ 384^{+ 14}_{- 41}$  \T \B \\
10.25--10.75     & 10.53        & 2.59  & 64  & 2       & $ 8.3^{+4.5}_{-4.1}$   & $ 739^{+ 27}_{- 26}$ \T \B \\
10.75--11.50     & 10.90        & 2.68  & 46  & 7   & $20.1^{+5.5}_{-5.3}$       & $1229^{+148}_{-139}$ \T \B \\
\hline
\end{tabular}
\end{table*}

\end{appendix}

\end{document}